\documentclass[
  journal=largetwo,
  manuscript=article-type,
  year=2020,
  volume=37,
]{cup-journal}

\usepackage{amsmath}
\usepackage{amssymb}
\usepackage[nopatch]{microtype}
\usepackage{booktabs}
\usepackage{gensymb} % degree synbol
\usepackage{xfrac}
\usepackage{aas-macros}

\usepackage{multirow}

\newcommand{\red}[1] {#1}
\newcommand{\old}[1] {#1}
\newcommand{\blue}[1] {#1}
\newcommand{\startobs} {2024-12-14}
\newcommand{\finishobs} {2025-03-31}
\newcommand{\minrate} {100}
\newcommand{\maxrate} {3000}

\newcommand{\totalgpsapprox} {$\sim$95000}

\newcommand{\totalfluenceidx} {$-3.17\pm0.02$} % Figure 5 in /home/msok/Desktop/EDA2/papers/2024/EDA2_FRBs/20250501_final_selfconsistent_fluence_distributions_PAPER.odt

\newcommand{\totalfluenceidxmp} {$-3.13\pm0.02$} % Figure 2 in /home/msok/Desktop/EDA2/papers/2024/EDA2_FRBs/20250501_final_selfconsistent_fluence_distributions_PAPER.odt
\newcommand{\totalfluenceidxip} {$-3.59\pm0.06$} % % Figure 3 in /home/msok/Desktop/EDA2/papers/2024/EDA2_FRBs/20250501_final_selfconsistent_fluence_distributions_PAPER.odt

\newcommand{\SEFDI} {$\text{SEFD}_I$}

\title{100,000 Crab giant pulses at 215\,MHz detected with an SKA-Low prototype station}

\author{M.~Sokolowski}
\affiliation{International Centre for Radio Astronomy Research, Curtin University, Bentley, WA6102, Australia}
\email[M.~Sokolowski]{marcinsokolastro@gmail.com}
\author{P.~Kumar}
\affiliation{International Centre for Radio Astronomy Research, Curtin University, Bentley, WA6102, Australia}

\author{S.~Dhavali}
\affiliation{International Centre for Radio Astronomy Research, Curtin University, Bentley, WA6102, Australia}

\author{B.~W.~Meyers}
\affiliation{Australian SKA Regional Centre (AusSRC), Curtin University, Bentley, WA 6102, Australia}
\alsoaffiliation{International Centre for Radio Astronomy Research, Curtin University, Bentley, WA6102, Australia}

\author{N.~D.~R.~Bhat}
\affiliation{International Centre for Radio Astronomy Research, Curtin University, Bentley, WA6102, Australia}

\author{A.~Bera}
\affiliation{International Centre for Radio Astronomy Research, Curtin University, Bentley, WA6102, Australia}

\author{S.~McSweeney}
\affiliation{International Centre for Radio Astronomy Research, Curtin University, Bentley, WA6102, Australia}

 %% First letter not capped

\begin{document}

\begin{abstract}
% REMOVED :  This is caused by scattering lowering the peak flux density and translating to lower detection rates of peak flux density optimised software packages like PRESTO.
We report detection and analysis of the largest ever low-frequency sample of Crab giant pulses (GPs) detected in frequency band 200 -- 231.25\, MHz. In total about \totalgpsapprox \, GPs were detected, which, to our knowledge is the largest low-frequency sample  of Crab GPs presented in the literature. The observations were performed between \startobs \, and \finishobs \, with the Engineering Development Array 2, a prototype station of the low-frequency Square Kilometre Array telescope. The fluence distribution of GPs in the entire sample is very well characterised with a single power law N(F) $\propto$ F$^\alpha$, where $\alpha =$\totalfluenceidx \,for all GPs, and $\alpha_{MP} =$\,\totalfluenceidxmp \, and $\alpha_{IP} =$\totalfluenceidxip \,for GPs at the phases of the main pulse and \old{low-frequency interpulse} respectively. We do not observe flattening of the fluence distribution at the higher fluences. Although, the index of the power law fluence distribution remained approximately constant over the observing period, the normalisation of the distribution was strongly anti-correlated (coefficient $\approx -0.9$) with the scatter broadening time. \old{The timescale ($\sim$ weeks) of these variations indicates that intrinsic GP emission was modulated by the refractive scintillation as the signals propagated through the Crab Nebula and ISM.} As a result, the measured fluence distribution was augmented for lower ($\tau \approx$ 2 ms) and diminished for higher ($\tau \approx$ 5 ms) scatter broadening time $\tau$ causing the GP detection rate to vary between \maxrate \, and \minrate \, per hour respectively (the correlation coefficient $\approx$-0.9). \red{Furthermore, for the first time at low-frequencies we observe indications of positive correlation (correlation coefficient $\approx$0.7) between the scatter broadening time ($\tau$) and dispersion measure.} Our modelling favours the screen size $\sim10^{-5}$\,pc with mean electron density $\sim 400$e$^{-}$cm$^{-3}$ located within 100\,pc from the pulsar (Crab Nebula or Perseus arm of the Milky Way galaxy). The observed frequency scaling of the scattering broadening time $\beta \approx -3.6\pm0.1$ (where $\tau \propto \nu^{\beta}$) is in agreement with the previous measurements. The observed maximum spectral luminosities $\sim 10^{25}$\,erg/Hz/s approach those of the weakest pulses from some repeating Fast Radio Bursts (FRBs). \red{However, the distribution of pulse arrival times is consistent with a purely random Poisson process, and we do not find evidence of clustering.} Overall, our results agree with the current views that GPs from extra-galactic Crab-like pulsars can be responsible for some very weak repeating FRBs, but cannot explain the entire FRB population. Finally, these results demonstrate an enormous transient science potential of individual SKA-Low stations, which can be unlocked by milli-second all-sky imaging.
\end{abstract}

% eda2_frb_pipeline.png
% NEW : ~/Desktop/SKA/papers/2024/EDA2_FRBs/PAPER/PIPELINE_FLOWCHART
% OLD : re-do image ~/Desktop/EDA2/papers/2024/EDA2_FRBs/meetings/20250217
\begin{figure*}[hbt!]
\centering
\includegraphics[width=0.98\linewidth]{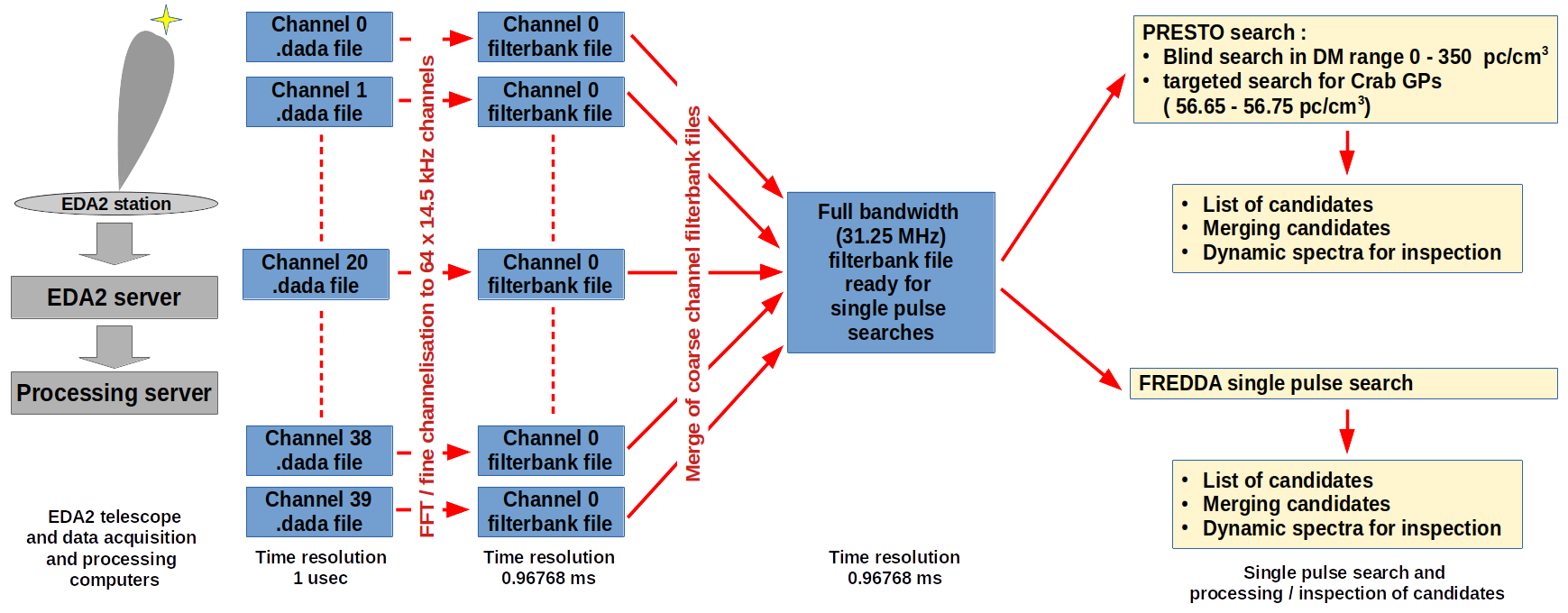}
\caption{Flowchart of the EDA2 FRB pipeline. From left to right: complex voltages from EDA2 real-time station beam at 1.08\,usec time resolution are recorded and saved on the data acquisition server (eda2-server on the left), with data from 40 coarse channels saved to separate 40 \textsc{dada} files. These files are copied to the data processing computer and processed off-line with a custom developed CPU/GPU spectrometer, which fine-channelised each coarse channel into 64 fine channels, time averaged resulting spectra to 0.96768\,ms time resolution and saved to 40 \textsc{filterbank} files. These files were merged into a single wide-band \textsc{filterbank} file, which was searched for single pulses with PRESTO \citep{2011ascl.soft07017R} and FREDDA \citep{2019ascl.soft06003B} software packages. The resulting single-pulse candidates were saved to text files for further processing and visual inspection of corresponding dynamic spectra.}
\label{fig_eda2_pipeline}
\end{figure*}

\section{Introduction}
\label{sec_introduction}
Pulsar B0531+21, aka Crab, is one of the first discovered pulsars \citep{1968Sci...162.1481S}. It is also one of the most interesting and longest studied pulsars capturing the imagination of the general public due to the magnificent Crab Nebula. It is a truly fascinating object and one of the ``living proofs'' that supernova (SN) explosions can lead to formation of a neutron star as the Crab pulsar and the surrounding nebula were formed in the core-collapse SN 1054 observed by medieval astronomers. As such, the Crab pulsar is considered to be a young pulsar with a period $P \approx 33.7$\,ms. It is located at the distance of approximately 1.9\,kpc \citep{2023ApJ...952..161L} and moving with a velocity of about 120\,km\,s$^{-1}$ \citep{2008ApJ...677.1201K,2023ApJ...952..161L}.

The Crab pulsar is one of a very few pulsars emitting the so-called giant pulses (GPs). \red{These are very short ($\sim\mu$s) bursts often exhibiting nanosecond substructure consisting of extremely narrow nano-shots (e.g., see \citet{2007ApJ...670..693H} or \citet{2010A&A...524A..60J}), which implies brightness temperatures reaching $10^{37}$\,K \citep{2003Natur.422..141H,2004ApJ...612..375C}.} Crab GPs were discovered shortly after the discovery of the pulsar \citep{1970Natur.226..529H,1970Natur.226...69S}. Their energies follow a power law distribution \citep{1972ApJ...175L..89A,2004ApJ...612..375C,2008ApJ...676.1200B,2015PASA...32....6O,2017ApJ...851...20M,2019MNRAS.490L..12B} in contrast to the exponential or log-normal distribution of normal pulses observed in other pulsars \citep{2012MNRAS.423.1351B}, which indicates different emission mechanisms. 

\red{Interestingly, Crab average pulse profiles are considered to be an average of a large ensemble of GPs (e.g. \citet{2021ApJ...920...38B,2021ApJ...915...65M}). Hence, even though the physical mechanisms behind the GP emission remain unexplained (see review by \citet{2016JPlPh..82c6302E}), they make the Crab pulsar a unique laboratory to study pulsar emission mechanisms (for example, see discussions in \citet{2007ApJ...670..693H,2010A&A...524A..60J,2021ApJ...920...38B}).}

\red{Crab's average pulse profile changes significantly as a function of frequency, and  8 components have been reported in literature (e.g. see \citet{2023ApJ...959..111L} for a detailed summary). At frequencies below 2\,GHz and at higher photon energies (from near infrared up to hard $\gamma$-rays) the average pulse profile consists of the main pulse (MP) with a precursor and a low-frequency interpulse (IP) at a phase offset of about 145\degree \,(about 13\,ms) from the MP. At these frequencies the average pulse profiles are consistent with the standard model that MPs and IPs originate from physically very similar regions on the opposite sides of the neutron star (NS). However, this picture becomes much more complicated at frequencies $\sim$2 -- 10\,GHz. Firstly, at around 2.7\,GHz the IP component disappears, while at higher frequencies ($\gtrsim$4\,GHz) the IP re-appears, but shifted in phase by about 20\degree \, from the low-frequency IP. Therefore, it is believed to be a different component, and referred to as high-frequency IP. Furthermore, two additional high-frequency components (HFC1 and HFC2) show up at pulse phases following the IP (see Figure 1 in \citet{2007ApJ...670..693H}). Finally, \citet{2007ApJ...670..693H} discovered that properties (temporal and spectral structures, DM and polarisation) of high-frequency GPs occurring at phases of MP and high-frequency IP differ significantly. This implies that the physical mechanisms and conditions in the emission region of the high-frequency IP are very different to those of the MP - in contrast to standard pulsar model where their are expected to be nearly the same just on the opposite sides of the NS. Similarly, the high-frequency IP and additional components (HFC1 and HFC2) are most likely unrelated to the low-frequency IP and originate from different regions and physical processes. All these relatively recent findings demonstrate that, despite being one of the longest studied pulsars,  Crab pulsar has still many mysteries, which can teach us a lot about the entire pulsar population. We note, however, that the micro and nano-second structure of Crab GPs cannot be observed at the low frequencies of the presented study due to the temporal scatter broadening caused by the ISM and Crab Nebula.}

After the discovery of fast radio bursts \citep[FRBs;][]{2007Sci...318..777L}, GPs from extra-galactic pulsars were hypothesised to be responsible for at least some types (i.e. repeating) of FRBs \citep{2016MNRAS.457..232C,2016MNRAS.462..941L}, and Crab GPs have been extensively studied in the context of FRBs (e.g.~\citet{2022NatAs...6..393N,2019ApJ...876L..23H}) as well as for testing this hypothesis (e.g. \citet{2024ApJ...973...87D,2019MNRAS.490L..12B,2017ApJ...851...20M}). Based on the energy distribution of Crab GPs and other characteristics (e.g. time between pulses), these studies have disfavoured the theory that all FRBs are caused by GPs from extra-galactic pulsars, and it is more likely that majority of FRBs are caused by objects like magnetars \citep{2019MNRAS.485.4091M,2021A&A...651A..63G}. However, some weaker FRBs, such as FRB 20200120E, could be caused by extra-galactic Crab-like pulsars (see ~\citet{2022NatAs...6..393N}). For the most recent review of the FRB field see \citet{2022A&ARv..30....2P}, while FRB models are collated at \url{https://frbtheorycat.org/}.

While the discovery and early Crab GP studies (e.g. \citet{1995ApJ...453..433L}) were performed at frequencies below 1\,GHz, the majority of later investigations were conducted at frequencies above 1\,GHz (e.g.~\citet{2024ApJ...973...87D,2019MNRAS.490L..12B,2019MNRAS.483.1224D,2008ApJ...676.1200B})\old{ up to even 15.1\,GHz (e.g.~\citet{2010A&A...524A..60J,2007ApJ...670..693H})}. On the high end of electromagnetic spectrum GP studies extend to $\gamma$-rays; see \citet{2021Univ....7..448A,} for the most recent review, and \citet{2021Sci...373..425L,2023ApJ...956...80A} for recent results at PeV energies.

The advent of new (LOFAR \citep{2013A&A...556A...2V}, MWA \citep{2013PASA...30....7T,2018PASA...35...33W}, LWA \citep{2013ITAP...61.2540E} etc.) and upcoming instruments, like low-frequency Square Kilometre Array telescope \citep[SKA-Low;][]{2009IEEEP..97.1482D}\footnote{\url{https://www.skatelescope.org/}} with large fields-of-view (FoVs), has begun to unlock the wealth of information available at these frequencies, and reinvigorated pulsar studies at frequencies below 400\,MHz. Several recent GP studies were performed down to 110\,MHz (e.g. \citet{2012A&A...538A...7K,2017ApJ...851...20M,2015ApJ...809...51O,2012ApJ...760...64M,2007ApJ...665..618B}). They addressed various aspects of Crab GPs ranging from monitoring plasma in the surrounding environment and along the line of sight (LoS) to the pulsar, through variations in dispersion measure (DM) and scatter broadening time to statistical properties of GPs, their emission mechanisms and similarities to FRBs.

\begin{figure*}[hbt!]
\centering

% /home/msok/Desktop/EDA2/papers/2024/EDA2_FRBs/20250327_calibrate_dynamic_spectrum_PAPER.odt
% mergedcand00233_start4370931_end4373816_NORM_CALIB_GIMPED.png
\includegraphics[width=0.42\linewidth]{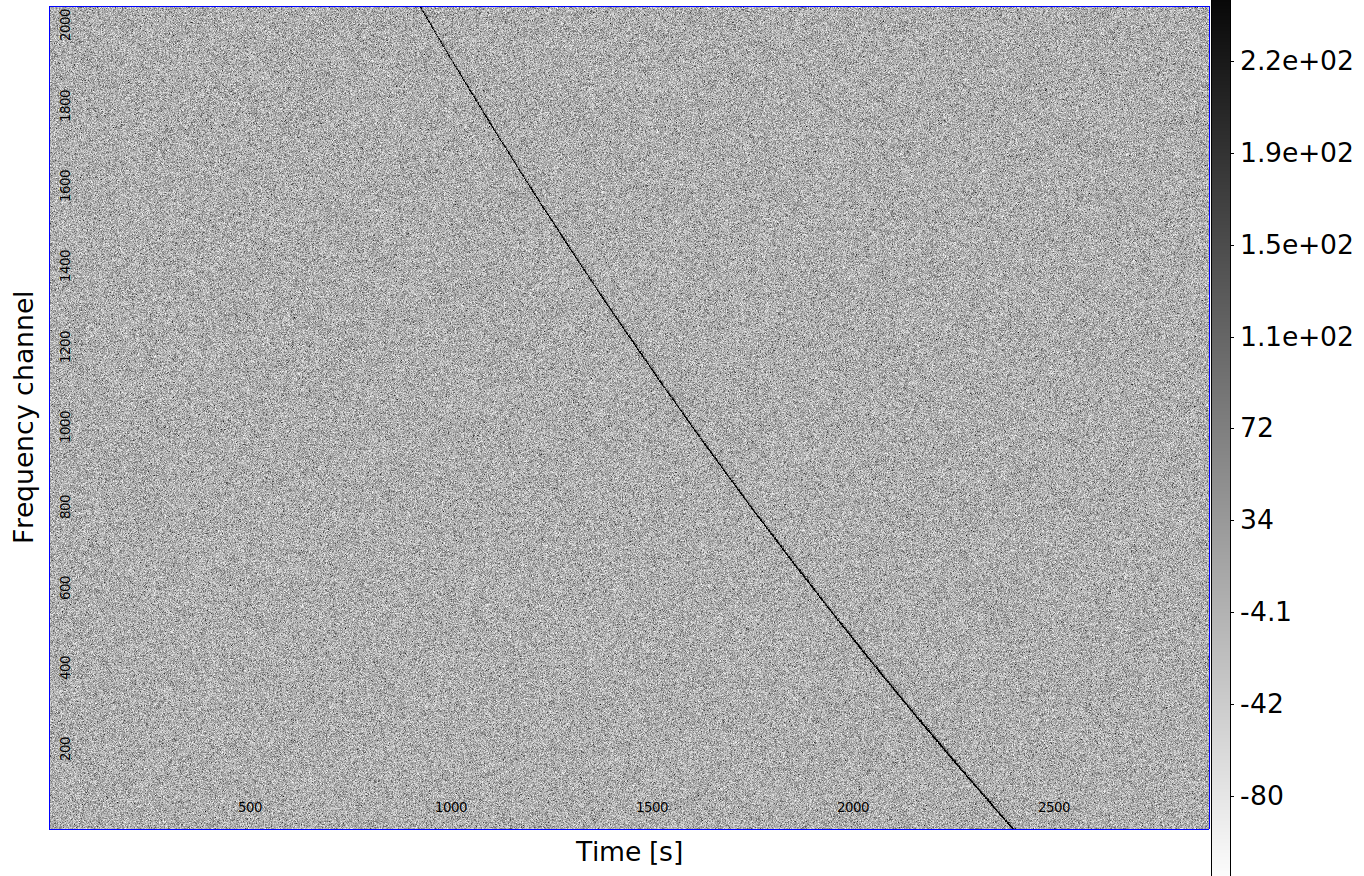}
% OLD :/home/msok/Desktop/EDA2/papers/2024/EDA2_FRBs/20250328_fitting_pulse_profile_PAPER.odt /home/msok/Desktop/SKA/papers/2024/EDA2_FRBs/PAPER/Pulse_profile_fit/images/FINAL/pulse00000_snr241_time4230sec_FIT_GAUSSxEXP.png
% NEW :
% /home/msok/Desktop/EDA2/papers/2024/EDA2_FRBs/20250328_fitting_pulse_profile_PAPER_FINAL.odt 
% /home/msok/Desktop/SKA/papers/2024/EDA2_FRBs/PAPER/Pulse_profile_fit/images/FINAL/pulse00000_snr241_time4230sec_FIT_GAUSSxEXP_LOG.png
% WARNING : see notes there PRESTO.txt max SNR ~ 241, my extract ~354 , FREDDA 372 -> using my as then I used my extractor for Fluence also it's close to FREDDA value
% MAX SNR/FLUX/FLUENCE : page 2 in /home/msok/Desktop/EDA2/papers/2024/EDA2_FRBs/20250328_fitting_pulse_profile_PAPER_FINAL.odt
% cat calibrated_pulses.txt
% # Time[sec] MAX_FLUX[Jy] FLUENCE[Jy ms] SNR SNR_FLUX[Jy] PRESTO_SNR PRESTO_FLUX[Jy]
% 4230.48600576 16606.10156250 76082.35932989 354.24627686 16606.10208299 241.59000000 11325.08202211
% 
% Gaussian FWHM = 2.355 sigma
% \includegraphics[width=0.53\linewidth]{pulse00000_snr241_time4230sec_FIT_GAUSSxEXP.png}
\includegraphics[width=0.53\linewidth]{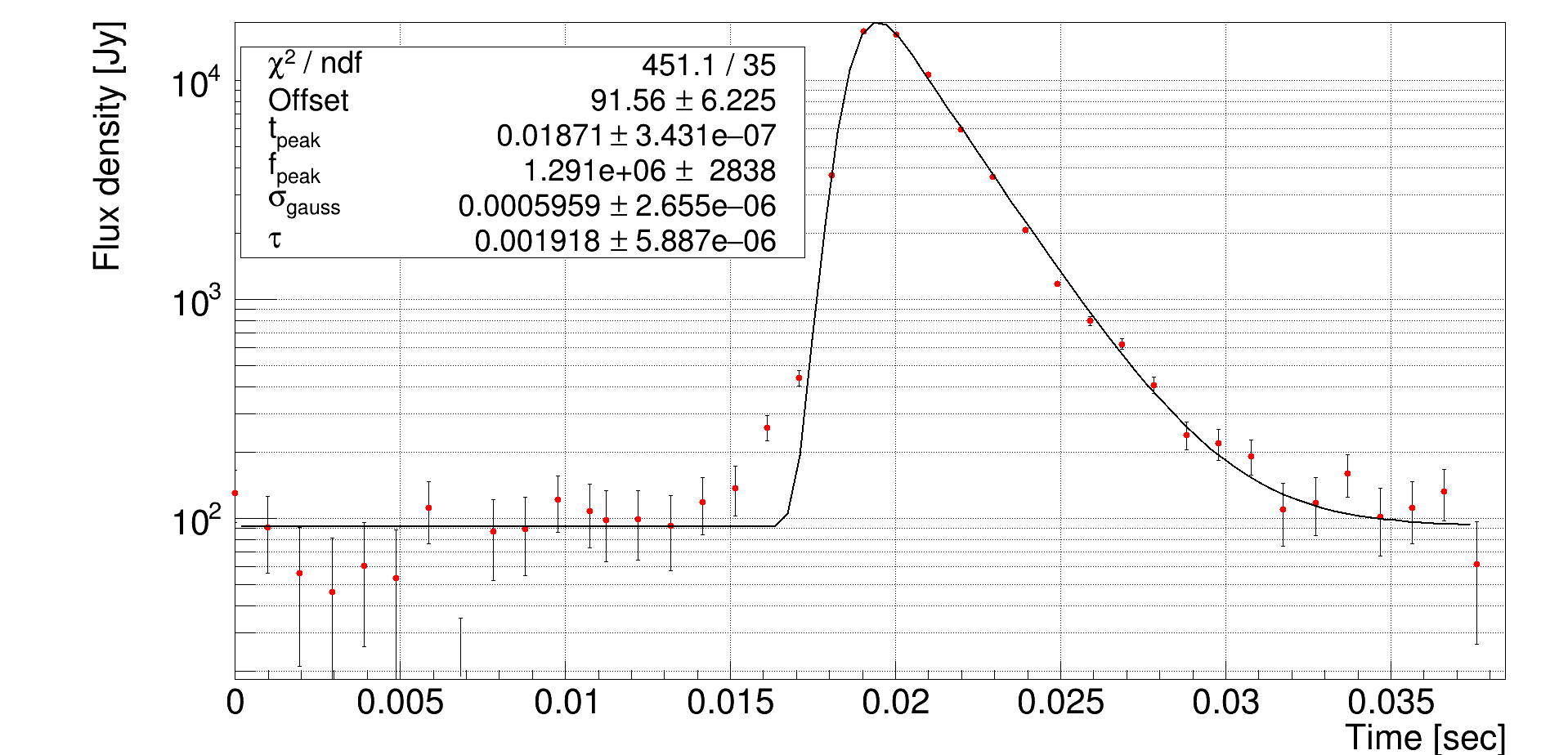}
\caption{The brightest Crab giant pulse detected with EDA2 (SNR $\approx$350). The corresponding flux density is $\approx$16.6\,kJy and fluence $\approx$76\,Jy\,ms. Left: dynamic spectrum calibrated in Jy. Right: profile after de-dispersion and averaging over the entire frequency band with a fitted function (Gaussian pulse with an exponential tail as per equation~\ref{eq_pulse_scatter_broadening} in ~\ref{appendix_one}). The fitted scatter broadening time $\tau = $1.918$\pm$0.006\,ms, and the FWHM of the Gaussian is FWHM$= 1.4 \pm 0.01$\,ms which is inline with the $\approx$1\,ms time resolution of the data.}
\label{fig_example_max_snr_gp}
\end{figure*}

This paper presents results from one such instrument, the Engineering Development Array version 2 \citep[EDA2;][]{2022JATIS...8a1010W} a prototype station of the SKA-Low. 
Despite modest sensitivity we were able to record the largest ever sample (\totalgpsapprox) of Crab giant pulses observed at SKA-Low frequencies, superseding even the largest ever sample of 30000 Crab GPs recorded by \citet{1995ApJ...453..433L}. The EDA2 was originally deployed as a risk mitigation and reference system to the official SKA-Low prototype station the Aperture Array Verification System 2 station \citep[AAVS2;][]{2022JATIS...8a1014M}. Real-time station beamforming and pulsar observing mode were commissioned at an early stage and already resulted in early SKA-Low pulsar science results \citep{2022PASA...39...42L,2024PASA...41...80L}. Both stations have also been used for all-sky transient searches \citep{2021PASA...38...23S}. 

Recently, a new pipeline, based on a code developed for the MWA by \citet{2024PASA...41...11S}, has been implemented and deployed for non-targeted (``blind'') searches and monitoring of repeating FRBs, magnetars, RRATs and pulsars with a single station beam of EDA2. In order to test and verify the pipeline, multiple observations of Crab pulsar were performed, and led to detection of thousands of GPs. These detections have confirmed that the station and the pipeline are capable of detecting highly dispersed pulses, and can be successful in searching for and/or monitoring FRBs, magnetars, RRATs or other fast transients. Especially, once the full FoV of the stations is unlocked by multi-beaming or high-time resolution all-sky imaging as proposed by \citet{2022aapr.confE...1S}. 

Moreover, the huge sample of Crab giant pulses turned out to be extremely valuable for scientific analysis. We have used this sample to perform statistical analysis of the Crab GPs, and as a demonstration of the potential of low-frequency telescopes for monitoring plasma in local environments and/or along the LoS towards pulsars, FRBs and other objects. Additionally, we have, once more, examined the possibility of extra-galactic Crab-like pulsars being FRB progenitors. 

% Especially, that the initial observation indicated sudden increase in the hourly rate of GPs. The GP rate varied significantly during the observing period and has been strongly correlated with scatter broadening times. This is expected as the single pulse software packages like \textsc{PRESTO}\,\citep{2011ascl.soft07017R} and \textsc{FREDDA}\,\citep{2019ascl.soft06003B}  are sensitive to peak flux density (not fluence). Larger scatter broadening times lead to reduction of peak flux density as the energy gets distributed over longer time interval (i.e. scattering tail), which leads to smaller number of GPs exceeding the detection threshold (typically 5 standard deviations of the noise, i.e. $5\sigma$).

This paper is structured as follows. In Section~\ref{sec_observations} we describe the EDA2 telescope and summarise the Crab observations performed for this paper. Section~\ref{sec_data_processing} describes the pipeline and data processing, while Section~\ref{sec_data_analysis} summarises data analysis. Results are presented in Section~\ref{sec_results}, and discussed in Section~\ref{sec_discussion}.
Finally, in Section~\ref{sec_summary} we summarise, make conclusive remarks and discuss future directions and prospects.

\section{EDA2 telescope and observations}
\label{sec_observations}

The data for this study were recorded between \startobs \, and \finishobs \, with the EDA2 station. The details of EDA2  can be found in the description paper \citep{2022JATIS...8a1010W}, and only a brief summary relevant to this project is provided here. The EDA2 has been a risk mitigation and a reference to the AAVS2 station, with the only major difference being the antenna design. The EDA2 comprises 256 MWA bow-tie dipoles, and was used as a reference to AAVS2 station (already de-commissioned) composed of 256 SKALA 4.1 antennas. Both instruments were located close to each other (some 500\,m apart) at the SKA-Low site, Inyarrimanha Ilgari Bundara observatory\footnote{\url{https://www.csiro.au/en/about/facilities-collections/mro}}. Analogue signals from individual antennas are digitised and the entire bandwidth (400\,MHz) channelised into 512 coarse channels in the backend Tile Processing Units \citep[TPMs;][]{2017JAI.....641014N}. One of the main tasks of TPMs is to form real-time station beams in 1.08\,usec time resolution over the entire SKA-Low observing bandwidth (50 -- 350\,MHz).

For this project, a single station beam and 40 coarse channels covering the frequency band 200.00 -- 231.25\,MHz (i.e. observing bandwidth BW$=$31.25\,MHz) were recorded. Observations with wider bandwidths are also possible, and the system has been confirmed to be stable at least up to a BW$\sim$100\,MHz. During each observing session 1\,h (all observations except 3) or 1.5\,h (3 observations) of data were recorded and centred on transit time of the Crab pulsar. The recorded station beam complex voltages were saved to \textsc{DADA} format files with complex voltages from each coarse channel saved to a separate file (i.e. 40 \textsc{DADA} files per observation in total).

\section{Data processing pipeline}
\label{sec_data_processing}

The recorded \textsc{DADA} files have been processed with \textsc{eda2frb} pipeline\footnote{\url{https://github.com/marcinsokolowski/eda2frb}} developed for this project (Fig.~\ref{fig_eda2_pipeline}). The pipeline has been based on the FRB search pipeline applied to MWA incoherent beam \citep{2024PASA...41...11S}\footnote{at\url{https://github.com/marcinsokolowski/mwafrb}}. It has been executed on a server computer installed at the observatory and consists of the following main steps. First, complex voltages in each coarse channel \textsc{DADA} file are fine channelised into 64 fine channels (approximately 14.5\,kHz each) by a custom-developed spectrometer using FFTW (CPU) or cuFFT (GPU) libraries (the choice is optional). The resulting spectra (in 69.12\,usec time resolution) are averaged to 0.96768\,ms time resolution, and saved to separate \textsc{filterbank} files. Then, 40 \textsc{filterbank} files from each coarse channel are merged into a single \textsc{filterbank} covering the entire observing bandwidth (31.25\,MHz). These steps are represented by blue boxes in Figure~\ref{fig_eda2_pipeline}.

% https://www.jb.man.ac.uk/pulsar/crab/crab2.txt
The single pulse search is performed on the resulting wideband \textsc{filterbank} file using both \textsc{PRESTO} \citep{2011ascl.soft07017R} and \textsc{FREDDA} \citep{2019ascl.soft06003B} software packages. One of the \textsc{PRESTO} executions was performed at Crab pulsar's DM=56.72 pc/cm$^3$ as measured during this period by the Jodrell Bank observatory\footnote{\url{https://www.jb.man.ac.uk/pulsar/crab/crab2.txt}}. Selected  FREDDA candidates were visually inspected, and an example of a GP with the highest signal-to-noise (SNR) ratio of $\approx$354 recorded in this campaign is shown in Figure~\ref{fig_example_max_snr_gp}. This is the brightest pulse in the sample with flux density of $\approx$16.6\,kJy and fluence $\approx$76\,Jy\,ms. We note that the pulse broadening effects were not removed in our analysis (e.g. via de-convolution, \citet{2003ApJ...584..782B}), which may lead to a factor of $\sim$2 underestimation of the intrinsic peak flux density (see discussion in ~\citet{2007ApJ...665..618B}).

\old{It is worth mentioning that some previous studies recorded low-frequency (600\,MHz) GPs upon receiving a trigger from high-frequency (1400\,MHz) detections (e.g. \citet{1999ApJ...517..460S}). However, in our study observations were performed at a single frequency (31.25\,MHz bandwidth centred on 215\,MHz), which was not strongly impacted by temporal pulse broadening, ionospheric scintillation or RFI \citep{2016ApJ...829...62E}. The moderate data rates ($\sim$ 0.5\,TB/h) enabled  us to record 1 -- 2\,h observations. Hence, we were able to identify GPs independently (without relying on any external information) by performing a single pulse search offline using the standard software packages (i.e. PRESTO and FREDDA).}

\section{Data analysis}
\label{sec_data_analysis}

% ??? 
\begin{figure*}[hbt!]
\centering
% NEW : DM_Tau_vs_DATE_20250502_SNR10.png
% after re-doing Tau fitting to all pulses with SNR>=10 :
%    /home/msok/Desktop/EDA2/papers/2024/EDA2_FRBs/20250502_DM_Tau_on_the_same_plot_PAPER.odt
% ~/Desktop/SKA/papers/2024/EDA2_FRBs/PAPER/DM_vs_TIME/20250502/images/FINAL/DM_Tau_vs_DATE_20250502_SNR10.png
% 
% OLD : DM_Tau_vs_DATE_GIMPED.png
% (Pearson correlation coefficient $\approx$0.6 and p-value = $3\cdot10^{-8}$)
% ~/Desktop/SKA/papers/2024/EDA2_FRBs/20250410_DM_Tau_on_the_same_plot.odt
% ~/Desktop/EDA2/papers/2024/EDA2_FRBs/PAPER/DM_vs_TIME/images/FINAL/DM_Tau_vs_DATE_GIMPED.png
\includegraphics[width=1.03\linewidth]{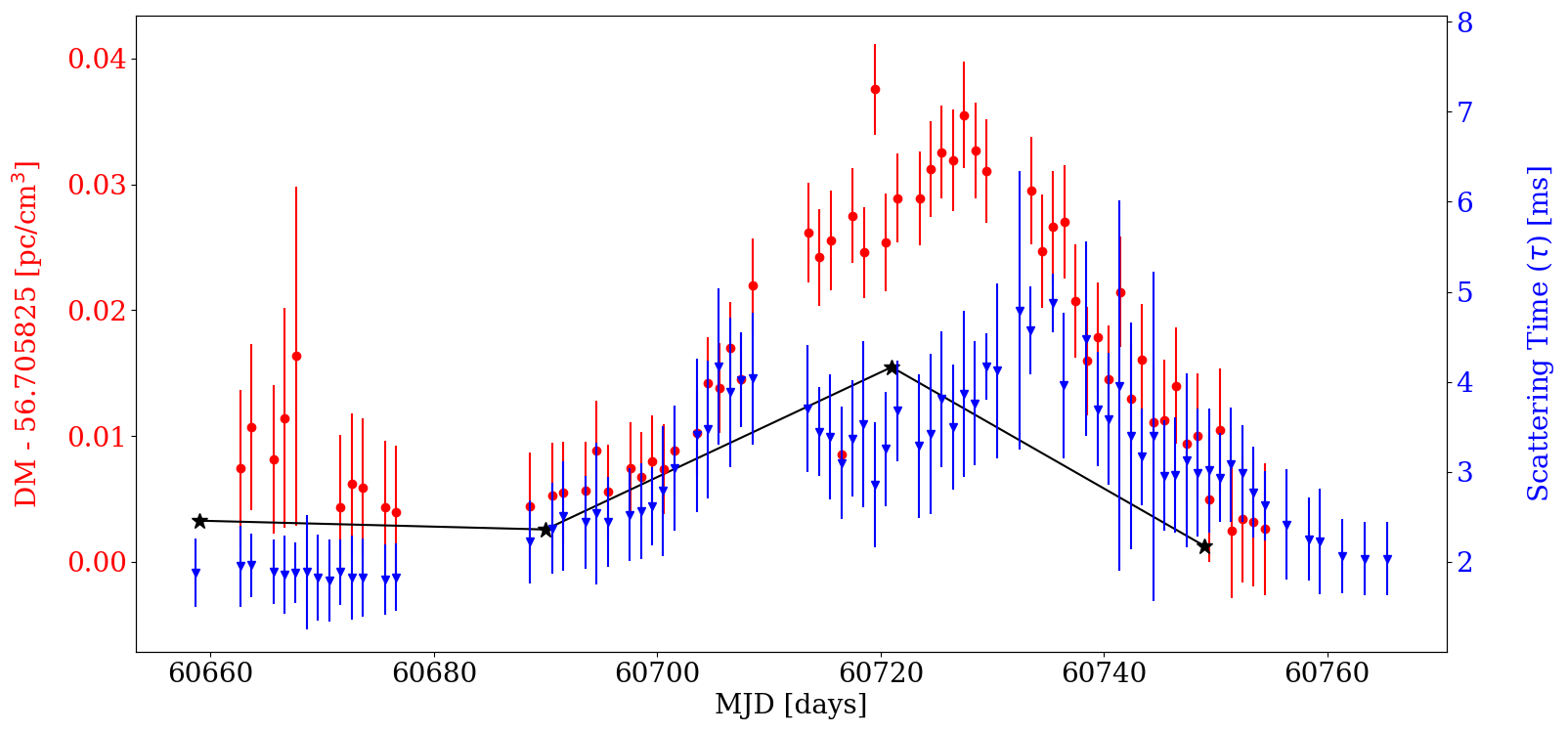}
\caption{DM (red circles) and scattering time (blue crosses) as a function of time during the time period \startobs \, to \finishobs. The black line connects the Jodrell Bank DM measurements (black stars) performed fortnightly. There is a good agreement between DM trends observed in this work and Jodrell Bank data. Furthermore, our data show strong correlation (Pearson correlation coefficient $\approx$0.7 and p-value = $10^{-8}$) between DM and scattering time ($\tau$). We note that the DM increase and correlation between DM and $\tau$ may be partially caused by changes of pulsar average profile due to scatter broadening. However, our additional verifications (see Section~\ref{subsec_scattering_and_dm}) and Jodrell Bank measurements show that there is at least $\approx$0.015 pc cm$^{-3}$ increase in DM. Hence, the results of our analysis are valid to within factor of 2.}
\label{fit_dm_and_scat_vs_time}
\end{figure*}

% /home/msok/Desktop/EDA2/papers/2024/EDA2_FRBs/20250411_Ngp_vs_Date_PAPER.odt
\begin{figure*}[hbt!]
\centering
% /home/msok/Desktop/SKA/papers/2024/EDA2_FRBs/PAPER/Ngps_vs_time/images/FINAL/Ngp_vs_time_PAPER.png
\includegraphics[width=0.98\linewidth]{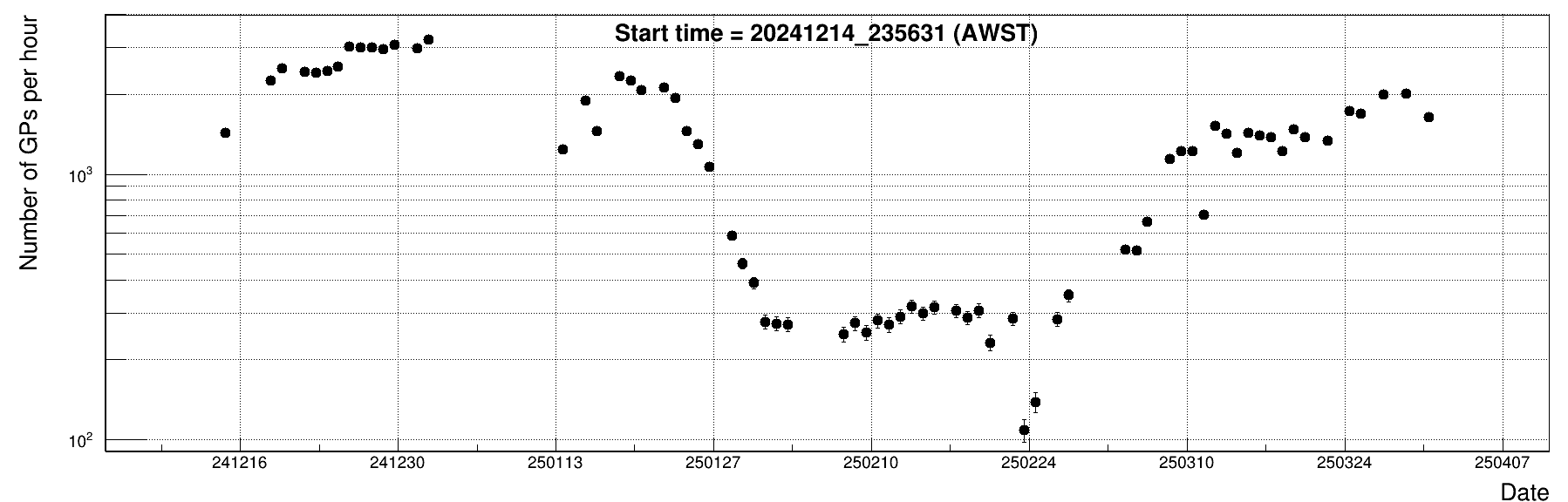}
\caption{Number of detected GPs as a function of time. The number of detected GPs is strongly correlated, in fact driven, by the scatter broadening (compare this figure with Figure~\ref{fit_dm_and_scat_vs_time} above and see also Figure~\ref{fig_ngp_vs_tau}). This comparison clearly shows that GP rate was very high (up to $\sim$3000\,h$^{-1}$) during in low-scattering conditions ($\tau \sim$2\,ms), and very low (down to $\sim$250\,h$^{-1}$) during high-scattering periods ($\tau \sim$4 - 5\,ms). }
\label{fig_ngp_vs_date}
\end{figure*}

% /home/msok/Desktop/SKA/papers/2024/EDA2_FRBs/PAPER/DM_vs_TIME/20250502/images/FINAL/Tau_vs_DeltaDM_FIT_PAPER.png
% Figure 2 in /home/msok/Desktop/EDA2/papers/2024/EDA2_FRBs/20250504_Tau_vs_DM_correlation_and_fit.odt
\begin{figure}[hbt!]
\centering
\includegraphics[width=0.98\linewidth]{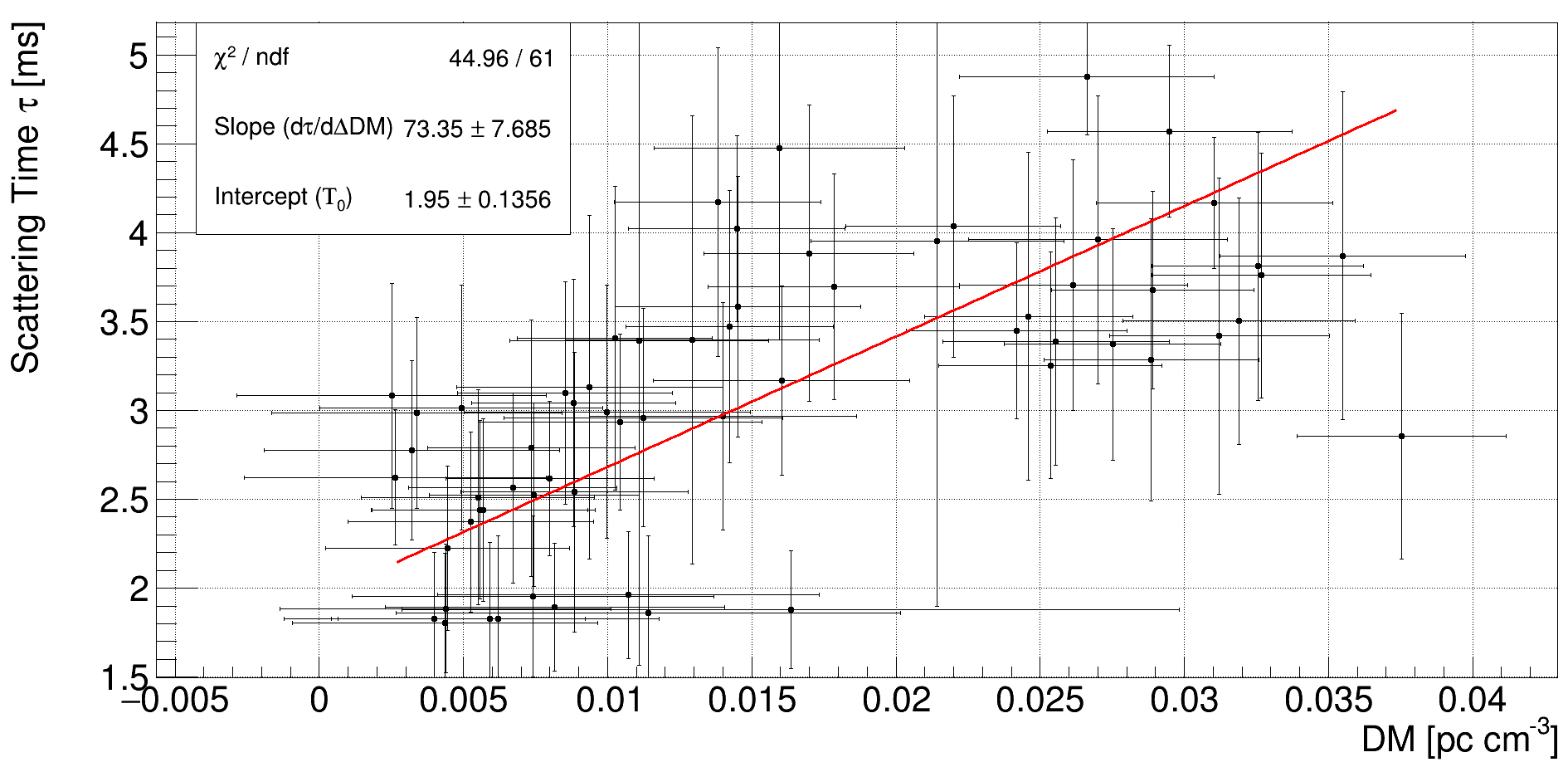}
\caption{Scatter broadening time ($\tau$) as a function of $\Delta$DM with a fitted linear function. The two quantities are highly correlated with correlation coefficient $\approx$0.7. \red{However, part of this correlation may be caused by the impact of scatter broadening on DM measurement based on pulsar timing analysis. The DM measurement based on maximising SNR of GPs yielded lower correlation coefficient $\sim$0.5. More robust DM measurement procedure in the presence of strong scattering is required to confirm this correlation.}}
\label{fig_tau_vs_deltadm}
\end{figure}

% OLD : /home/msok/Desktop/EDA2/papers/2024/EDA2_FRBs/20250411_Ngp_vs_Date_PAPER.odt
% NEW : /home/msok/Desktop/EDA2/papers/2024/EDA2_FRBs/20250401_modelling_number_of_GPs_PAPER.odt
\begin{figure}[hbt!]
\centering
% OLD :/home/msok/Desktop/SKA/papers/2024/EDA2_FRBs/PAPER/Ngps_vs_time/images/FINAL/Ngps_vs_taufit_PAPER.png
% \includegraphics[width=0.98\linewidth]{Ngps_vs_taufit_PAPER.png}
% OLD :   /home/msok/Desktop/SKA/papers/2024/EDA2_FRBs/PAPER/Ngps_vs_time/PULSE_MODEL/NOISE/images/FINAL/ngps_vs_tau_5000Jyms_SIMUL_PAPER.png
% \includegraphics[width=0.98\linewidth]{ngps_vs_tau_5000Jyms_SIMUL_PAPER.png}
% 
%
%%%%%%%%%%%%%%%%%%%%%%%%%%%%%%%%%%%%%%%%%%%%%%%%% NEW :
% NEW (2025-05-04)  : page 3 Figure 1 in /home/msok/Desktop/SKA/papers/2024/EDA2_FRBs/20250503_modelling_number_of_GPs_PAPER.odt
% ~/Desktop/EDA2/papers/2024/EDA2_FRBs/PAPER/MODELLING_Ngps/images/FINAL$ Ngps_vs_tau_PAPER_GIMPED.png
\includegraphics[width=0.98\linewidth]{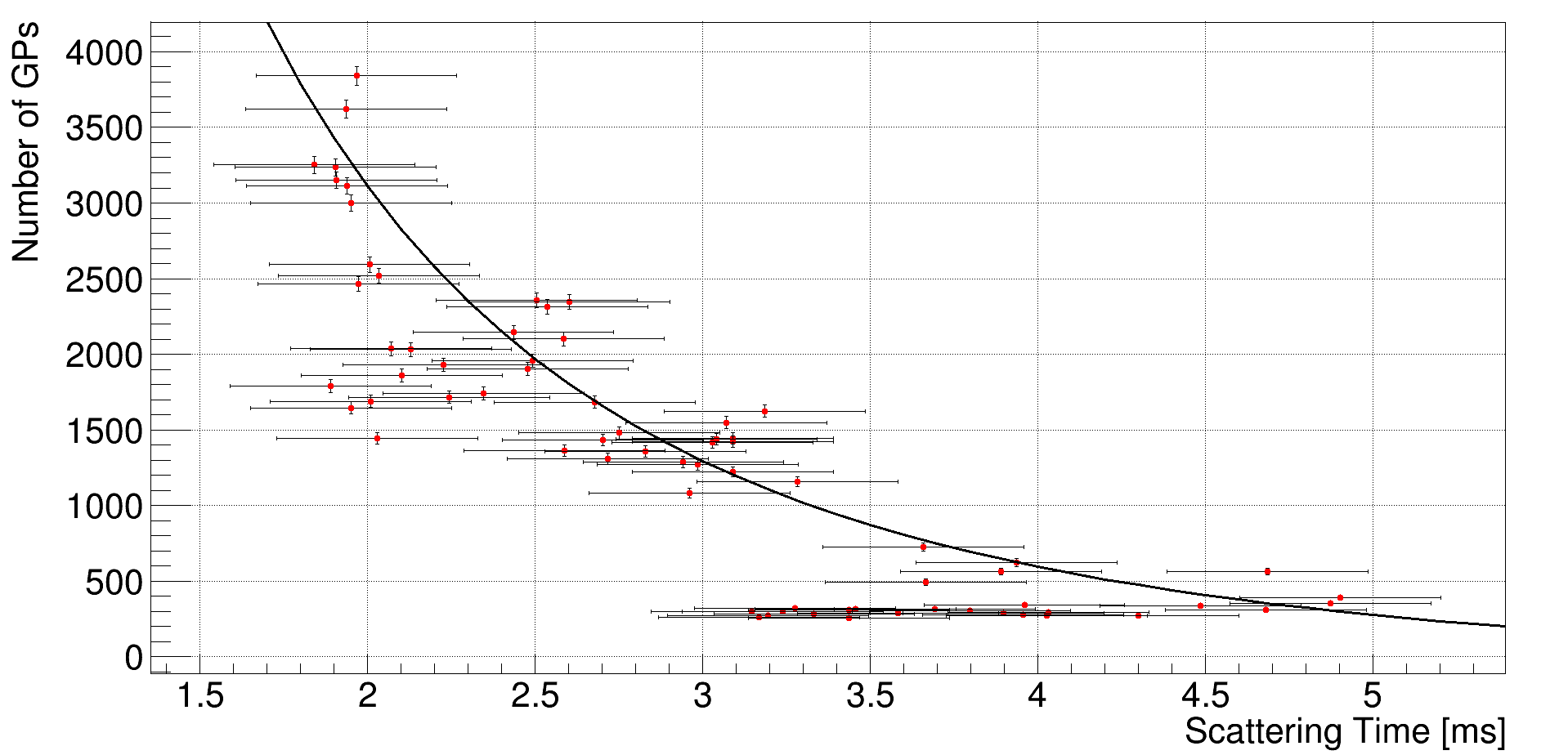}
\caption{Number of detected GPs is strongly anti-correlated with the scattering time, with Pearson and Spearman correlation coefficients $\approx -0.9$ and p-value $\sim 10^{-22}$. The black curve is the model described in Section~\ref{subsec_modelling_dm}.}
\label{fig_ngp_vs_tau}
\end{figure}

\subsection{Post-processing of PRESTO output}
\label{subsec_postprocessing_presto}

The output from PRESTO \textsc{single\_pulse\_search.py} consists of three types of files: \textsc{dat} binary files contain de-dispersed time series as a series of float values, \textsc{inf} text files contain metadata, and \textsc{singlepulse} files contain a list of identified single pulse candidates (including arrival time, DM etc.). 
The following additional steps were performed on PRESTO output in preparation for the later analysis:

\begin{itemize}
% Twin pulses search : /home/msok/Desktop/EDA2/papers/2024/EDA2_FRBs/20250509_looking_for_twin_GPs_MPs_vs_IPs.odt
% ~/Desktop/SKA/papers/2024/EDA2_FRBs/references/CRAB_TWIN_GPs/MS
% 2010A&A...515A..36K
\item \textbf{Merging candidates} - a single GP may result in multiple candidates with different arrival times, DMs and SNRs. Therefore, in order to avoid double-counting, the candidates were merged into a single candidate with maximum SNR within a 45\,ms time window as highly scattering-broadened pulses can extend up to this duration. In principle, we could have missed two GPs within one pulsar period ($\approx$33\,ms), but the results of merging were visually inspected and such cases have never been found. In fact, so far, the so-called double (or ``twin'') GPs, from MP and IP within the same pulsar period, have only been reported in L-band \citep{2010A&A...515A..36K}. \red{Also, recently, \citet{2023ApJ...959..111L} reported detections of clusters of micro-bursts (even up to 6) within a single pulse window ($\lesssim$1\,ms), but such instances cannot be identified in our data due to scatter broadening at our frequencies and $\sim$1\,ms time resolution.}

\red{In order to find double-GPs, we performed a dedicated search in} every recorded de-dispersed time series, and did not find any instance of two pulses separated by less than a pulsar period ($\approx$33\,ms). Hence, we are confident that the merging procedure does not affect the statistical analysis of fluence and flux density distributions, or any other parts of the presented analysis.

\vspace{0.1cm}

\item \textbf{Extraction of de-dispersed time series} - The de-dispersed time series for each dataset were extracted from the PRESTO binary files (\textsc{dat} files) generated by the de-dedispersion at DM=56.72 pc/cm$^3$. First, a running median and standard deviation (derived from the interquartile range) were calculated. The running median value was then subtracted as a baseline and the resulting time series was divided by the running standard deviation, which converted the uncalibrated flux units of the timeseries into SNR units. The resulting SNR-timeseries was flux-calibrated in Jy using the procedure described in Section~\ref{subsec_calibration}, and flux density, fluence and SNR of each PRESTO candidate (after merging) were saved to the output text file. The timeseries extraction code is available at \url{https://github.com/marcinsokolowski/presto_tools}.
\end{itemize}

\subsection{Flux Density Calibration}
\label{subsec_calibration}

The de-dispersed time series converted to SNR units (Section~\ref{subsec_postprocessing_presto}) and SNRs reported in the \textsc{.singlepulse} log files were flux-calibrated using equation \ref{eq_calibration}. If $f_{data}$ is the uncalibrated flux density measured in the data (de-dispersed timeseries), the calibrated flux density $f_{cal}$ was calculated as:

\begin{equation}
f_{cal} = f_{data} \frac{\sigma_{sim}}{\sigma_{data}} = \text{SNR}_{data} \sigma_{sim},
\label{eq_calibration}
\end{equation}

where $\sigma_{sim}$ and $\sigma_{data}$ are standard deviation of the noise as expected from the simulation and measured in data respectively, and $\sigma_{sim} / \sigma_{data}$ ratio is the calibration factor (or inverse of the gain). The expected standard deviation of the noise ($\sigma_{sim}$) can be calculated from the radiometer equation:

\begin{equation}
\sigma_{sim} = \frac{\text{SEFD}_I}{\sqrt{\Delta \nu \Delta t}},
\end{equation}

where $\Delta \nu$ and $\Delta t$ are observing bandwidth and integration time respectively. $\text{SEFD}_I$ is the System Equivalent Flux Density (SEFD) for Stokes I polarisation (hence $n_{pol}=1$ and this factor is not included under the square root). For each dataset, \SEFDI \, was calculated using the SKA-Low-sensitivity calculator \citep{2022PASA...39...15S}\footnote{\url{https://github.com/marcinsokolowski/station_beam}}, which provides very accurate SEFD values as verified by extensive tests during the EDA2 and AAVS2 commissioning and verification stages \citep{2022JATIS...8a1010W,2021ecap.confE...1S}. The mean value during the Crab pulsar transit at elevation $\approx$41\degree was \SEFDI$\approx7930$\,Jy (for reference EDA2 SEFD at zenith is $\approx$2500\,Jy). Hence, the expected noise was $\sigma_{sim} \approx46$\,Jy for the observing parameters $\Delta \nu=31.25 \cdot 10^6$\,Hz (=31.25\,MHz), $\Delta t=0.00096768$\,sec ($\approx$1\,ms). 

The flux-calibrated timeseries (in Jy) was used to calculate calibrated fluence (in units of Jy\,ms) as a sum (i.e. discrete integral) of the flux density under the pulse. For each GP reported by PRESTO (after merging), the fluence $F$ was calculated as the following sum:

\begin{equation}
F = \Delta t \sum_{i} f_{i},
\label{eq_fluence_sum}
\end{equation}

where $\Delta t$ is the time bin width ($\approx$1\,ms), $f_{i}$ is the flux density in the bin $i$, and the sum extends over the positive flux density bins surrounding the peak flux. The sum includes all positive flux density values before and after the flux density peak. Hence, the summation stops when the first negative value is encountered before and after the peak flux density bin, which indicates that the flux density reached the level consistent with noise. Negative values are present because of the subtraction of the running median as a baseline.

After calibration of each dataset (typically one hour observation per day), all parameters (arrival time, SNR, calibrated flux density, calibrated fluence etc.) describing every GP detected by PRESTO (after merging) were saved to a single text file (\textsc{calibrated\_pulses.txt}). This file was a starting point for further statistical analysis. The scripts used for this processing are publicly available\footnote{\url{https://github.com/marcinsokolowski/crab_frb_paper}}.

\section{Results}
\label{sec_results}

% Completeness thresholds are self-consistent now see /home/msok/Desktop/SKA/papers/2024/EDA2_FRBs/20250405_tests_of_completness_threshold.odt
The text file \textsc{calibrated\_pulses.txt} served as a starting point for further analysis.
The completeness threshold was verified to be SNR$\ge$10, which at low scattering conditions ($\tau \sim$2\,ms) corresponded to flux density threshold $f_{min} \approx $\,500\,Jy, and fluence threshold $F_{min}=$1400 Jy\,ms. A scatter-plot (not included here) of all GP SNR vs. fluence confirmed that F$\ge$1400 Jy\,ms above this value there are no GPs approaching the PRESTO threshold SNR$\lesssim$5 which would not be detected (even during high scattering period when peak flux density may be reduced). Hence, no GPs with fluence F$\ge$1400 Jy\,ms were missed (i.e. the sample is complete).

\subsection{Fluence distribution}
\label{subsec_fluence_distrib}

The GPs from the main pulse (MP) and interpulse (IP) were separated by calculating GP phase (using time of arrival and pulsar period) and matching the phase of MP and GP in the average profile. The fluence distributions of GPs associated with MP and IP were separately plotted and fitted with a power law (Figure~\ref{fig_fluence_distributions}):

\begin{equation}
N(f) = N_{ref} \left( \frac{f}{f_{ref}} \right)^\alpha, 
\label{eq_power_law}
\end{equation}

where $N_{ref}$ is the normalisation corresponding to number of counts at the reference value $f_{ref}$ (1000\,Jy\,ms for the fluence and 1000\,Jy for the flux density distributions respectively). Therefore, the fitted value of $N_{ref}$ can be interpreted as an hourly rate of GPs with these respective values of fluence or flux density.

Approximately 85\% of the detected GPs occur at phase of the MP (the remaining 15\% at phase with IP), which is in very good agreement with 86\% measured at similar frequencies with MWA and $\sim$84\% at 732\,MHz \citep{2017ApJ...851...20M}. It is also very close to $\sim$87\% at 1200\,MHz \citep{2012ApJ...760...64M}.
% Bradley's paper  ~94\% at 3100\,MHz therein 

\begin{figure}[hbt!]
\centering
% /media/msok/5508b34c-040a-4dce-a8ff-2c4510a5d1a3/eda2/crab_paper/PAPER/images/FINAL/Nref_vs_tau_PAPER.png
% /home/msok/Desktop/SKA/papers/2024/EDA2_FRBs/20250502_refitting_fluence_distributions.odt
\includegraphics[width=0.98\linewidth]{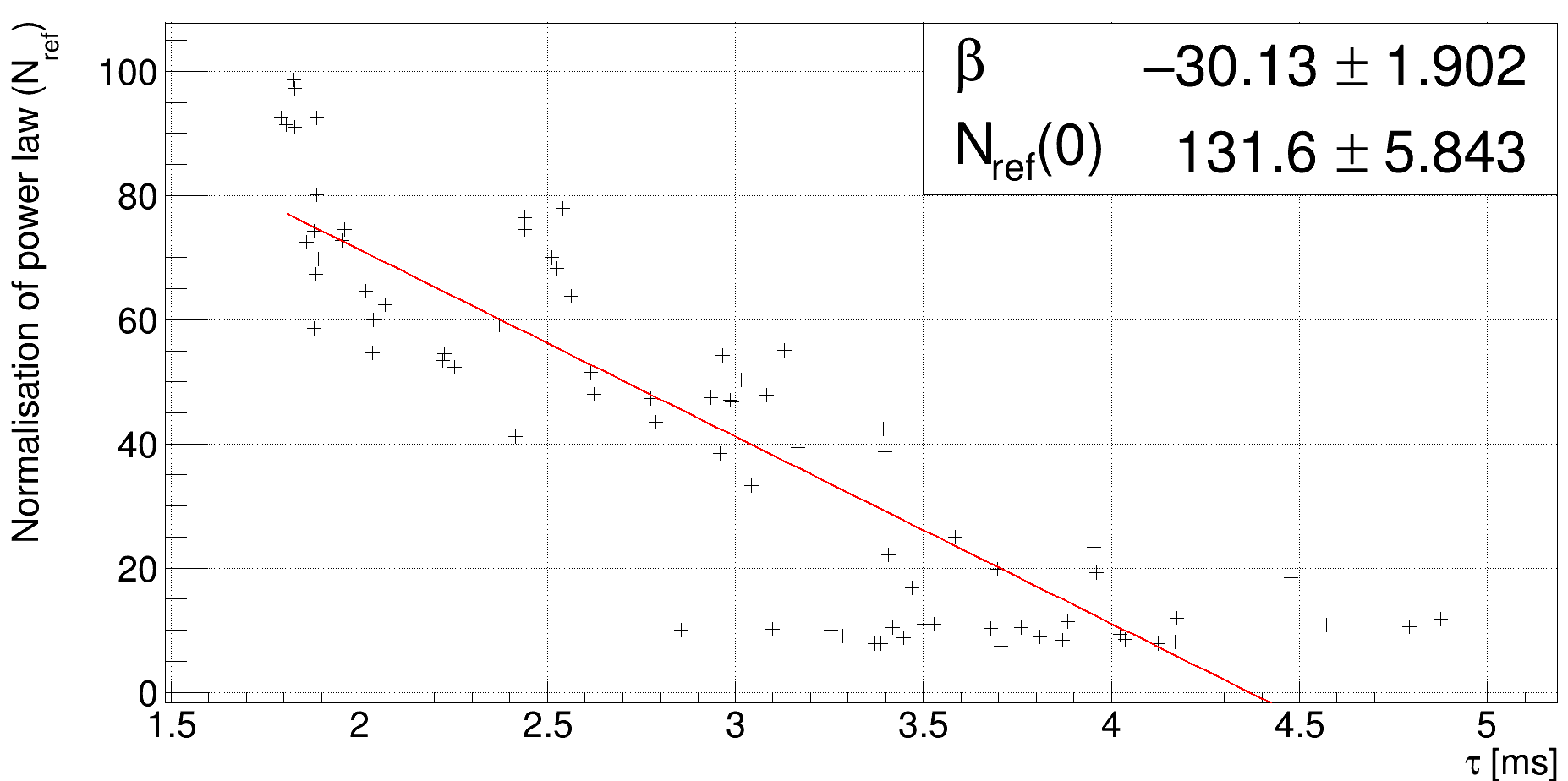}
\caption{The normalisation of the fluence distribution ($N_{ref}$ in equation~\ref{eq_power_law}) was strongly anti-correlated with the scattering time ($\tau$), with Pearson and Spearman correlation coefficients $\approx -0.9$ and p-value $\sim 10^{-22}$. The red line is the fitted linear function $N_{ref}(\tau) = N_{ref}(0) + \beta \tau$. This correlation and the observed timescale strongly 
suggest that the observed variability was caused by the refractive scintillation.}
% \todo{Update if I end up using [0]/(x-[1])+[2] or similar.}
\label{fig_plnorm_vs_tau}
\end{figure}

% /home/msok/Desktop/SKA/papers/2024/EDA2_FRBs/20250502_refitting_fluence_distributions.odt
The power law index $\alpha$ fitted to separate datasets was  approximately constant as a function of time ($\alpha = -3.50 \pm 0.2$). On the other hand, the fitted power law normalisation ($N_{ref}$) was strongly anti-correlated (coefficient $\approx$0.9 and p-value $\sim 10^{-25}$) with the scatter-broadening time ($\tau$). Hence, $N_{ref}$ was higher when the $\tau$ was lower and vice versa (Fig.~\ref{fig_plnorm_vs_tau}). This resulted in the fluence distribution ``floating'' up and down under the low ($\tau \sim$2\,ms) and high ($\tau \sim$5\,ms) scattering conditions respectively. Consequently, the number of detected GPs was also strongly correlated (coefficient $\approx$0.9) with the scatter broadening time (compare Figures~\ref{fig_ngp_vs_tau} and ~\ref{fit_dm_and_scat_vs_time}) and increased and decreased when the scattering time was lower and higher respectively (Fig.~\ref{fig_ngp_vs_tau}).

Interestingly, our results confirm (probably yet another time) that PRESTO is a very good ``fluence'' detector as higher scattering times did not compromise the detection efficiency. This is based on the observation that the turn-over fluence $F_{peak}$, which is a fluence below which the number of detected GPs starts to decrease due to missing fainter GPs, remains approximately constant for various scatter broadening times. A weak correlation between $F_{peak}$ and $\tau$ was indeed observed, indicating that as expected, higher scattering time reduces peak flux density resulting in fewer fainter pulses detected (i.e. $F_{peak}$ is higher for higher $\tau$ values). However, this correlation is much weaker than could be expected for purely peak flux based detection software, while PRESTO implements boxcar convolution which makes it much closer to a perfect ``fluence detector''.

\subsection{Scatter broadening and DM variations}
\label{subsec_scattering_and_dm}

The scattering time was measured using GPs with SNR$\ge$10. Each of these high-SNR pulses was fitted as a Gaussian pulse with an exponential tail (equation~\ref{eq_pulse_scatter_broadening} in ~\ref{appendix_one}), which yielded scatter broadening time $\tau$ (e.g.~Figure~\ref{fig_example_max_snr_gp}). For each dataset all SNR$\ge$10 GPs were fitted, and the mean and standard deviation of the resulting scattering time distribution provided the mean scattering time and its error for a particular dataset (i.e. date). 

Scattering time and DM as a function of time are shown in Figure~\ref{fit_dm_and_scat_vs_time}, where DM measured by Jodrell Bank \citep{1993MNRAS.265.1003L}\footnote{\url{http://www.jb.man.ac.uk/~pulsar/crab/crab2.txt}} is also plotted as a reference. A systematic constant offset between DM measured in this work and Jodrell Bank measurements is noticeable (see also Figure~\ref{fig_dm_timing_vs_gps} and discussion in ~\ref{appendix_two}). Especially, at MJDs between the start and 60680 days, before the increase in scattering time begun, which could affect our low-frequency timing-based DM measurements. Such DM offsets are not uncommon for measurements performed at different frequencies (this work at 215\,MHz  and Jodrell Bank in L-band), and may result from chromatic DM (see, for example\citet{2019ApJ...882..133K,2016ApJ...817...16C}). However, this does not affect our analysis.

% details in  or minimising rise time of the leading edge
The DM was measured using standard pulsar timing performed with the \textsc{TEMPO2} software package using frequency dependent pulse profiles. These measurements may be impacted (overestimate DM) by the changes of the average pulsar profile due to scattering - mainly by shifting the peak and changing the shape of the average pulse profile. Therefore, as a cross-check the DM was also measured by fitting profiles of individual GPs and finding DM maximising their peak flux density (i.e. SNR). The comparison of the DM measured with this method and timing is shown in Fig.~\ref{fig_dm_timing_vs_gps} in ~\ref{appendix_two}, where we also discuss other tested methods of measuring DM from individual GPs. The DM values measured from individual GPs are similar to the results of the timing analysis, and confirm that to the first order the latter are correct and the increase in DM is of physical origin (not an artefact of analysis). Although, during high scatter broadening period (between MJD 60700 and 60740), the increase in DM may be slightly lower than shown by the red filled circles in Figure~\ref{fit_dm_and_scat_vs_time}, the DM values are lower bound by the well-established Jodrell Bank measurements (red stars). This ensures that our further analysis is valid to within a factor of 2 (in the worst case). 

\red{There are indications of positive correlation} between DM and scattering time (Fig.~\ref{fig_tau_vs_deltadm}), and as the scattering time gets higher DM also increases and vice versa (the Pearson correlation coefficient $\approx$0.7 and the p-value $\approx 10^{-8}$). However, we note that this correlation may be slightly exaggerated by the earlier-mentioned impact of scattering on timing-based DM measurements. A similarly strong correlation of scatter broadening time and DM (correlation coefficient 0.56$\pm$0.01) based on normal pulses was observed in Crab pulsar at 610 MHz by \citet{2018MNRAS.479.4216M}, and at 350\,MHz by \citet{2019MNRAS.483.1224D}. Hence, our measurements are consistent with the earlier observations of scatter broadening and DM correlations at higher frequencies. \red{Nevertheless, we note that more data and robust DM measurements methods are required to confirm the correlation in the presence of strong scatter broadening.}

% Figure 1 in /home/msok/Desktop/EDA2/papers/2024/EDA2_FRBs/20250428_lag_correlation_of_DM_and_Tau.odt
% TEST in /home/msok/Desktop/EDA2/papers/2024/EDA2_FRBs/20250428_lag_correlation_of_DM_and_Tau_TEST.odt
The lag correlation of DM and scatter broadening time variations yielded a lag consistent with zero (to within a daily cadence of our observations). This is consistent with the results reported by \citet{2018MNRAS.479.4216M}, but significantly differs from the 30\,days lag observed by \citet{2008A&A...483...13K}. It seems that the DM/$\tau$ increase event observed by \citet{2008A&A...483...13K} was caused by a plasma screen with a different structure and/or geometry of the Earth, pulsar, screen.  

Finally, we have also compared scatter broadening times with earlier observations. Thus, the measured scatter broadening times were scaled to 610\,MHz with a scaling factor $(610/215.625)^{-3.5}$ as measured for the Crab pulsar \citep{2019ApJ...874..179K,2007ApJ...665..618B,2017ApJ...851...20M}. The resulting distribution is very similar and only slightly shifted towards lower scatter-broadening times with respect to the distribution in Figure 3 in \citet{2018MNRAS.479.4216M}. This indicates that the scattering conditions were likely similar during both events, and possibly caused by similar plasma structures.

\begin{figure*}[hbt!]
\centering
% /home/msok/Desktop/EDA2/papers/2024/EDA2_FRBs/20250408_flux_density_distrib_PAPER.odt
% /media/msok/5508b34c-040a-4dce-a8ff-2c4510a5d1a3/eda2/crab_full_analysis_final/merged/PAPER_PLOTS/Flux_density_distrib/images/FINAL/flux_density_distribution_paper.png
\includegraphics[width=0.45\linewidth]{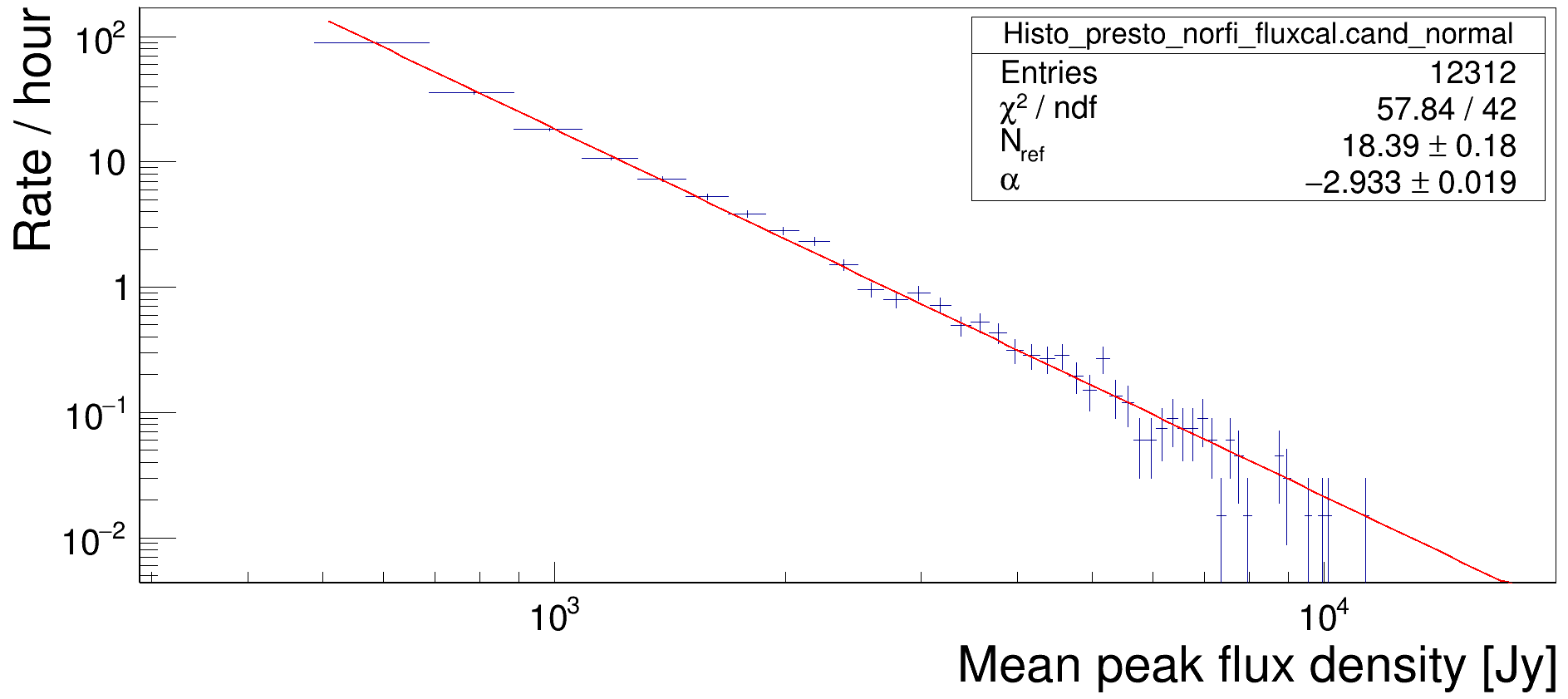}
% /media/msok/5508b34c-040a-4dce-a8ff-2c4510a5d1a3/eda2/crab_full_analysis_final/merged/PAPER_PLOTS/Flux_density_distrib/images/FINAL/spectral_luminosity_distribution_paper.png
\includegraphics[width=0.45\linewidth]{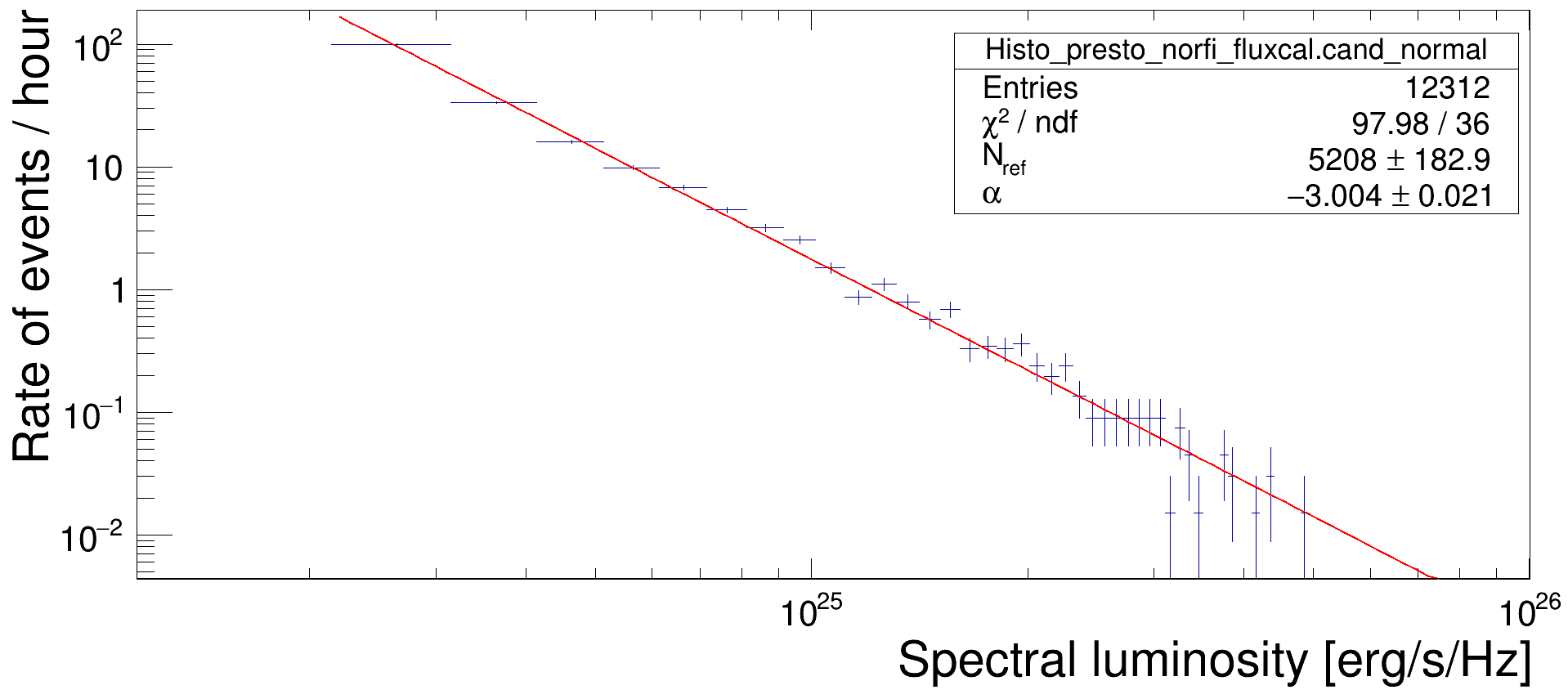}
\caption{Left: distribution of peak flux density, Right: distribution of spectral luminosity calculated according to equation~\ref{eq_spectral_luminosity}. Both were fitted with a power law (eq.~\ref{eq_power_law}) above the completeness threshold of SNR$\ge$10 and the resulting indexes $\alpha$ (i.e. slopes) are consistent within the errors, which is expected since the spectral luminosity was calculated from the peak flux density. It can also be seen from the right plot that the maximum spectral luminosity giant pulses in the sample approach $L= 5 \cdot 10^{25}$ erg/s/Hz.}
\label{fig_flux_density_distribution}
\end{figure*}

% \begin{figure*}[hbt!]
% \centering
% /home/msok/Desktop/EDA2/papers/2024/EDA2_FRBs/20250408_flux_density_distrib_PAPER.odt
% /media/msok/5508b34c-040a-4dce-a8ff-2c4510a5d1a3/eda2/crab_full_analysis_final/merged/PAPER_PLOTS/Fluence_distrib/images/FINAL/mp_fluence_distribution_paper.png
% \includegraphics[width=0.45\linewidth]{mp_fluence_distribution_paper.png}
% /media/msok/5508b34c-040a-4dce-a8ff-2c4510a5d1a3/eda2/crab_full_analysis_final/merged/PAPER_PLOTS/Fluence_distrib/images/FINAL/ip_fluence_distribution_paper.png 
% \includegraphics[width=0.45\linewidth]{ip_fluence_distribution_paper.png}
% \caption{Fluence distributions of giant pulses during the main pulse (left) and interpulse (right). Similarly, to earlier distributions the index of power law is close to $-3$. The power law fitted to GPs from the interpulse is slightly steeper ($\alpha_{ip}$ -3.4$\pm$0.1) than from the main pulse ($\alpha_{mp}$ -2.98$\pm$0.03). The fraction of GPs from MP is approximately 85\%, which is in a very good agreement with other studies.}
% \label{fig_fluence_distributions}
% \end{figure*}

\begin{figure*}[hbt!]
\centering
% OLD was not self-consistent in terms of binning with my fitting ROOT SCRIPT 
% see SOLUTION IN : /home/msok/Desktop/EDA2/papers/2024/EDA2_FRBs/20250501_Consistency_of_number_of_10Jyms_pulses_per_hour.odt
% NEW Figure is described in :
% NEW : page 4 of /home/msok/Desktop/EDA2/papers/2024/EDA2_FRBs/20250501_final_selfconsistent_fluence_distributions_PAPER.odt
% NEW : ~/Desktop/SKA/papers/2024/EDA2_FRBs/PAPER/FLUENCE_DISTRIB/images/FINAL/MP_IP_distribution_SELFCONSIST_GIMPED.png
% 
% OLD : Figure 1 in /home/msok/Desktop/EDA2/papers/2024/EDA2_FRBs/20250411_rate_of_occurance_PAPER-ONE-PLOT.odt
% OLD : /home/msok/Desktop/SKA/papers/2024/EDA2_FRBs/PAPER/Rate_of_occurance/MP_IP_distribution.png
\includegraphics[width=0.95\linewidth]{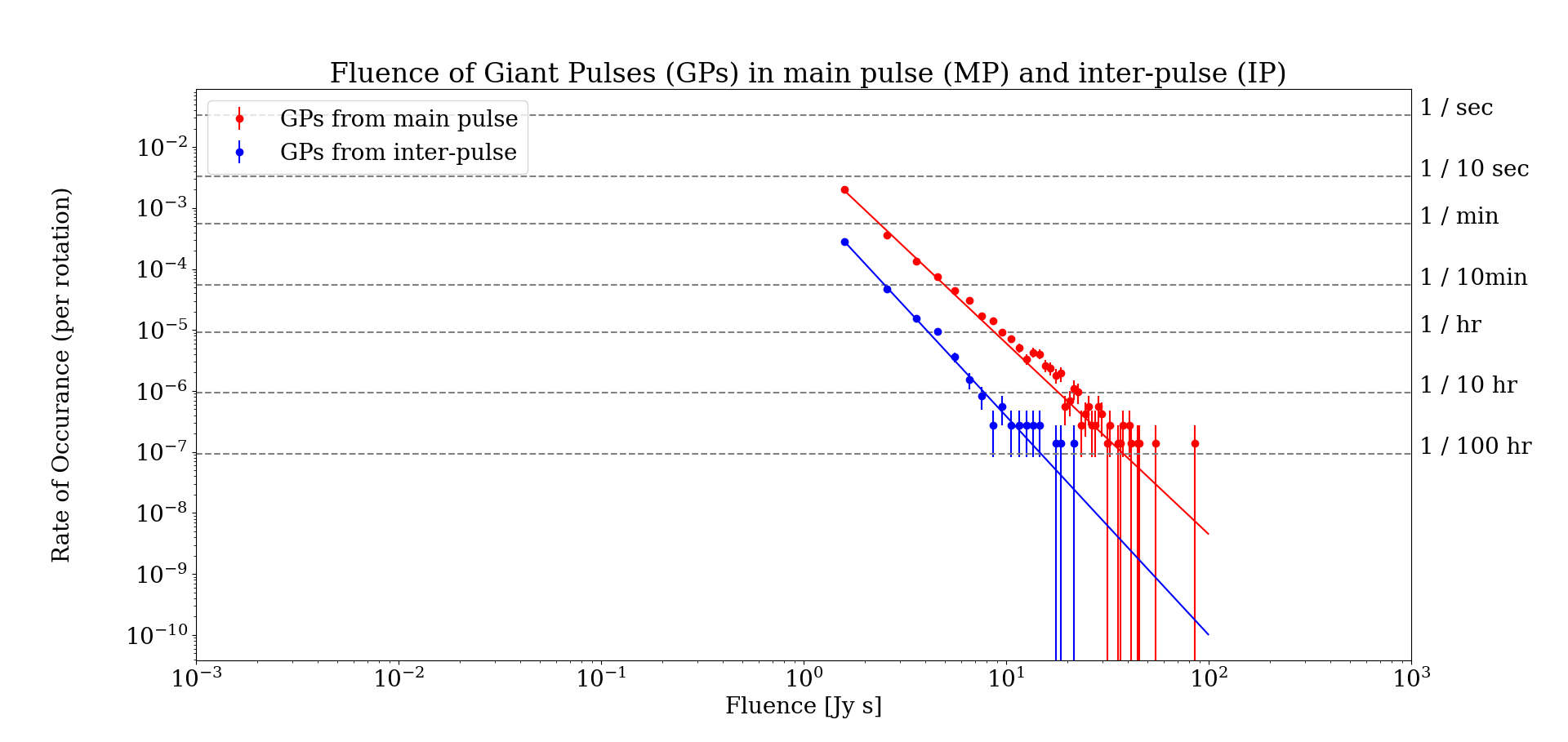}
\caption{Fluence distributions of giant pulses which occurred at phase of the main pulse (red points) and interpulse (blue points) with fitted power law function (lines in the corresponding colours). Similarly, to earlier distributions the index of power law fitted to all the GPs is \totalfluenceidx. The power law fitted to GPs at the phase of interpulses is slightly steeper ($\alpha_{ip}$ \totalfluenceidxip) than from the main pulse ($\alpha_{mp}$ \totalfluenceidxmp). The fraction of GPs from MP is approximately 85\%, which is similar to earlier studies. The units, scales, and ranges were selected to make easy comparisons with Figure 6 in \citet{2017ApJ...851...20M}. The fluence corresponding to 1 GP per hour is approximately 8500\,Jy\,ms (i.e. 8.5\,Jy\,s in the figure) for the fluence bin width $\Delta F \approx$1000\,Jy\,ms (i.e. 1\,Jy\,s in the figure).}
\label{fig_fluence_distributions}
\end{figure*}

\subsection{Number of GPs vs. scatter broadening}
\label{subsec_num_gps_vs_time}

One of the most interesting aspects originally noticed during the first observations in December 2024, were large variations in the numbers of detected GPs between 1-hour observations on different days (i.e. the rate of detectable GPs varied). During the first night of observations (\startobs) about 1400 GPs were detected by PRESTO, while on 2024-12-25 it reached 3000 to drop down to around 250 in the middle of February 2025 (Fig.~\ref{fig_ngp_vs_date}). Interestingly, it started to increase again at the end of February 2025 to reach 2000 at the end of March (on 2025-03-27). The number of detected GPs was strongly correlated with the scatter broadening time (compare Figures~\ref{fit_dm_and_scat_vs_time} and ~\ref{fig_ngp_vs_date}, and see correlation in Figure~\ref{fig_ngp_vs_tau}).

% ~/Desktop/SKA/papers/2024/EDA2_FRBs/PAPER/DIAGRAM/
\begin{figure}[hbt!]
\centering
\includegraphics[width=0.95\linewidth]{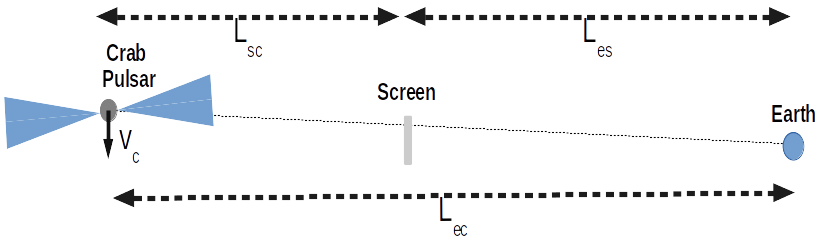}
\caption{The model of a plasma blob passing through the line of sight (LoS) between the observer on Earth and the pulsar. The motion of the blob can either be real (e.g. expansion of the Crab Nebula), or apparent due to the motion of the pulsar with velocity 120\,km/s \citep{2008ApJ...677.1201K,2023ApJ...952..161L}. The following distances defined in the image are used in the main text: $L_{es}$ is distance between the Earth and the screen, $L_{sc}$ is the distance between the screen and the Crab pulsar, and $L_{ec}$ is the distance between the Earth and the Crab pulsar (known to be about 2\,kpc).}
\label{fig_plasma_blob_model}
\end{figure}

% Modelling
The number of detected GPs as a function of scattering time $\tau$ has been modelled (black curve in Figure~\ref{fig_ngp_vs_tau}) using the scatter broadening time dependent fluence distribution (the fitted linear function in Figure~\ref{fig_plnorm_vs_tau}), and functional form of pulse profile. Equation ~\ref{eq_pulse_scatter_broadening} in ~\ref{appendix_one} was used to model giant pulse profiles as a function of time for different values of scatter broadening time $\tau$ with other parameters fixed ($t_{peak} = 0$, $f_{offset} = 0$, and $\sigma_t = 0.8$\,ms). 
As discussed in Section~\ref{subsec_num_gps_vs_time} the fluence distribution of GPs depends on scatter broadening time. Hence, we used the linear dependence as fitted in Figure~\ref{fig_plnorm_vs_tau} to calculate normalisation of the fluence distribution $N_{ref}$ for a specific $\tau$. Then the expected number of GPs for particular value of $\tau$ was calculated as an integral of the fluence distribution above the minimal fluence $F_{min}$. This minimal fluence $F_{min}$ was calculated as the fluence corresponding to the SNR threshold ($T_{snr}$) used by the detection software PRESTO (SNR=5, i.e. peak flux density = 5$\sigma_n \approx$230\,Jy\,ms). The number of GPs as a function of $\tau$ and the results of our modelling are shown as a black curve in Figure~\ref{fig_ngp_vs_tau}. We note that this is only an approximation, because as discussed earlier by using convolution with boxcar function, PRESTO effectively implements a fluence detector, and this has not been reflected in our model. However, even this simple model reproduces the observed dependence very well. This gives us confidence that the variations in number of detected GPs can be explained by the up and down variations of the fluence distribution caused by refractive scintillation, which is also supported by the $\sim$weeks timescale of these variations. Similar, variations in GP rates were observed by \citet{1995ApJ...453..433L} in their largest ever recorded sample of GPs, which they also attributed to propagation effects. The explanation in terms of refractive scintillation is discussed in Section~\ref{subsec_refractive_scintillation}.

\subsection{Flux density, SNR and spectral density distributions}
\label{subsec_fluxdensity_distribution}

Since, flux density was calculated from SNR using equation~\ref{eq_calibration}, these two quantities are related by a linear relation and are expected to follow the same distribution. The left image in Figure~\ref{fig_flux_density_distribution} shows the distribution of the peak flux density of GPs in the entire sample. The distribution is well described by a power law fitted above the completeness threshold of $f_{min}=$\,500\,Jy. The resulting power law index is $\alpha_{flux} = 2.93 \pm 0.02$, while the fit to the SNR distribution yielded a very similar value of $\alpha_{SNR} = 3.07 \pm 0.02$.

Using the calibrated peak flux density and the distance to the pulsar  D=2\,kpc \citep{2023ApJ...952..161L,2008ApJ...677.1201K}, the peak spectral luminosity was calculated according to the following equation:

\begin{equation}
L= 4 \pi D^2 f_{peak} = 4.32 \cdot 10^{21} \text{erg/s/Hz},
\label{eq_spectral_luminosity}
\end{equation}

where $f_{peak}$ is the peak flux density of the pulse. The resulting distribution is shown in the right panel of Figure~\ref{fig_flux_density_distribution}. The maximum values are  L$\lesssim  5 \cdot 10^{25}$erg/s/Hz, which is consistent with the Crab spectral luminosities reported in literature (e.g. compare with Fig. 3 in \citet{2022NatAs...6..393N} or Fig. 1 in \citet{2018NatAs...2..865K}). These maximum spectral luminosities are a 2 -- 3 orders of magnitude lower than the weakest repeating FRBs. However, assuming that there is enough energy in the system, the wait time for $10^{28}$erg/s/Hz pulse is relatively short (of the order of 50,000\,years). Hence, extra-galactic Crab-like pulsars may be responsible for weak repeating FRBs from the local Universe (see discussion in Section~\ref{subsec_spectral_luminosities}). However, the observed spectral luminosities are more than 10 orders of magnitude lower than typical or brightest FRBs. Hence, it is highly unlikely that such FRBs can be produced by objects similar to Crab pulsar (see discussion in Section~\ref{subsec_frb_context}) as there is not enough energy available. Additionally, based on our spectral luminosity distribution, the waiting time for such bright pulses ($\sim 10^{34}$erg/s/Hz) is more than 10 orders of magnitude longer than the Hubble time.
% /home/msok/Desktop/EDA2/papers/2024/EDA2_FRBs/20250531_number_of_galaxies.odt

\section{Discussion}
\label{sec_discussion}

In this section, we discuss modelling of the observed variability of DM, and scatter broadening time $\tau$ as a function of time, and its physical interpretation. We also interpret our results in the context of FRBs and re-examine the possibility of extra-galactic Crab-like pulsars being FRB progenitors.

\subsection{Modelling DM and $\tau$ variations}
\label{subsec_modelling_dm_and_tau}

% SIZE from Duration /home/msok/Desktop/EDA2/papers/2024/EDA2_FRBs/20250503_blob_size_from_the_duration.odt
The DM and $\tau$ variations have been modelled as a plasma screen (or a ``blob'') passing through the line of sight between the Earth and the Crab pulsar (Figure.~\ref{fig_plasma_blob_model}). 

% /home/msok/Desktop/EDA2/papers/2024/EDA2_FRBs/20250522_simple_Tau_modelling_FINAL.odt
% ~/Desktop/SKA/papers/2024/EDA2_FRBs/references/Ramesh_Bhat/0437/MS$ Gupta_1995ApJ...451..717G_MS.pdf
We consider two distances between Earth and the screen ($L_{es}$): (i) standard case midway (1000\,pc) between the Earth and pulsar, and (ii) close to the pulsar (some $\sim$1900\,pc from Earth) and consistent with the Crab Nebula or more broadly Perseus arm of the galaxy (explanation will be provided in Section~\ref{subsec_refractive_scintillation}). These two cases correspond to the ratio x=$L_{es}/L_{sc}$ (Fig.~\ref{fig_plasma_blob_model}) equal to 1 for (i) and $\sim$20 for (ii) respectively. For x=1 the velocity of the scintillation pattern with respect to observer on Earth is approximately equal to the velocity of the pulsar (eq.~\ref{eq_velocities}), while it can be higher for $x>1$ due to ``lever-arm'' argument (see equation~\ref{eq_velocities} or discussion in~\cite{1995ApJ...451..717G}).

% 2025-05-22 - replaced with discussion of mean delta-DM etc (later):
% As can be seen from Figures~\ref{fig_dm_model_fit} and Figures~\ref{fig_tau_model_fit} the entire event takes around $\Delta T=$75\,days, which corresponds to the blob size of $\sim 10^{-5}$\,pc for 100\,km/s and $\sim 10^{-6}$\,pc for 1000\,km/s. All parameters resulting from modelling are summarised in Table~\ref{tab_comparison}.

\subsubsection{Modelling DM variations}
\label{subsec_modelling_dm}
% ~/Desktop/EDA2/papers/2024/EDA2_FRBs/
% FINAL : 20250423_fitting_DM_and_Tau_variations_PAPER_FINAL.odt
% OLD - with BUG in normalisation : 20250418_scatter_broadening_time_modelling.odt
% 20250419_fitting_DM_and_Tau_variations.odt

% NEW 2025-05-05 : 
% Figure 1 in 20250505_DM_modelling_multiple_cases.odt
% Figure 1 in 20250423_fitting_DM_and_Tau_variations_PAPER_FINAL_SIMPLE.odt
% OLD VERSION (non-independent fitting):
% Figure 3 in /home/msok/Desktop/EDA2/papers/2024/EDA2_FRBs/20250423_fitting_DM_and_Tau_variations_PAPER.odt
% /home/msok/Desktop/SKA/papers/2024/EDA2_FRBs/PAPER/DM_vs_TIME/images/FINAL/DM_Tau_vs_DATE_FIT-5par_PAPER.png
\begin{figure}[hbt!]
\centering
\includegraphics[width=0.98\linewidth]{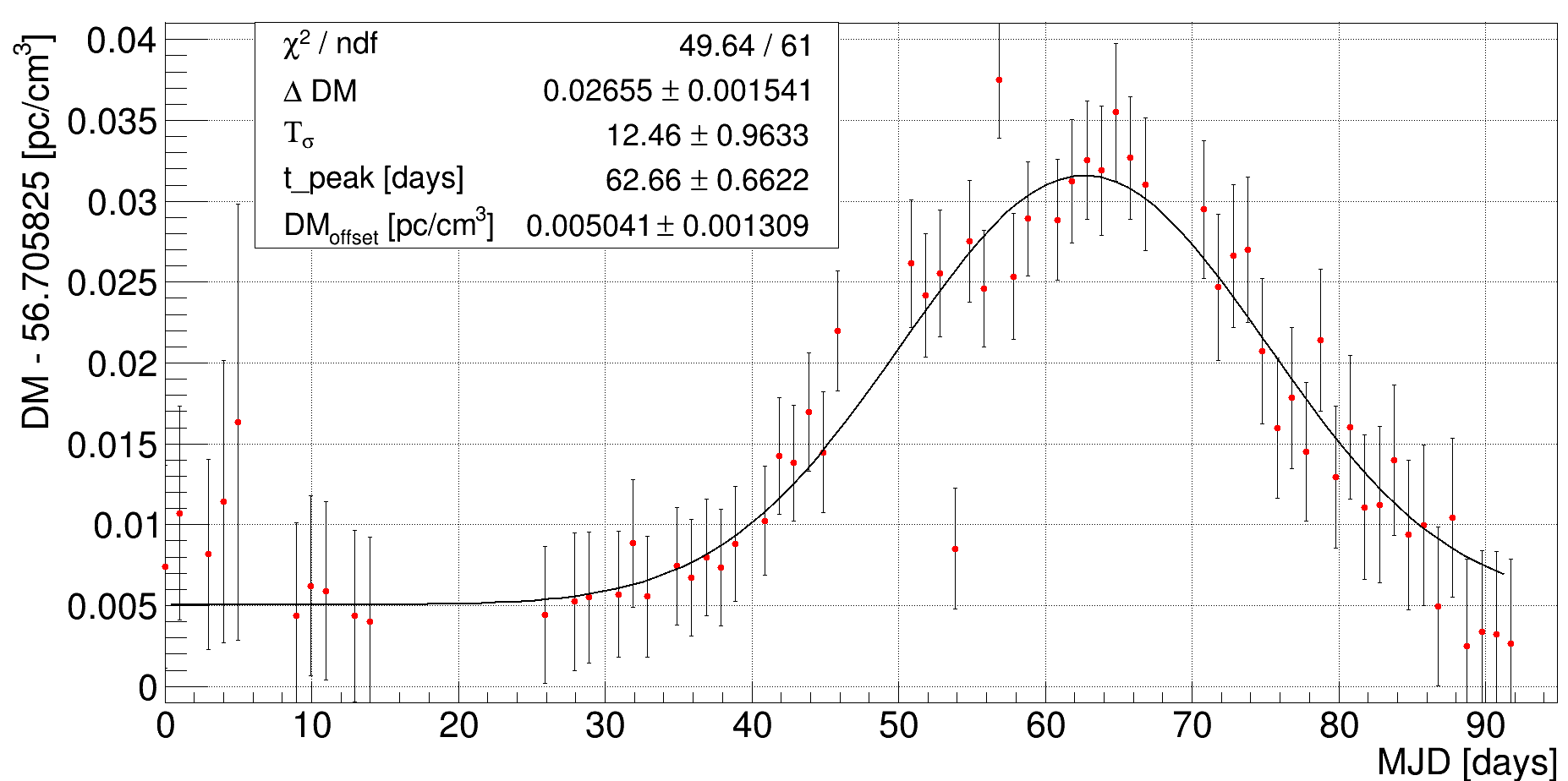}
\caption{DM vs. time fitted with equation~\ref{eq_gaussian_blob} describing a spherical plasma cloud passing transversely through the LoS between the pulsar and Earth. The plasma cloud is modelled as a sphere of Gaussian distributed electron density.}
\label{fig_dm_model_fit}
\end{figure}

% The electron density of the blob was modelled as a Gaussian distribution of the width ($\sigma_{blob}$), peak electron density ($n_{e}^{peak}$) and peak DM time ($t_{peak}$) free parameters of the fit in the following equation:
The DM variations shown in Figure~\ref{fit_dm_and_scat_vs_time} have been modelled as a sphere of electron plasma passing transversely through the LoS between the pulsar and Earth (Fig.~\ref{fig_plasma_blob_model}) using the following equation:

\begin{equation}
DM(t) = DM_{offset} + \Delta DM \cdot e^{-(t-t_{peak})^2/2\Tau_{\sigma}^2},
\label{eq_gaussian_blob}
\end{equation}

where the free parameters of the fit were $DM_{offset}$ additional small DM offset, $\Delta DM$ the amplitude of the DM peak, $\Tau_{\sigma}$ the timescale of the transition of the plasma sphere through the LoS. These free parameters are independent, and they can be expressed as the following physical parameters: $\Delta DM = \sqrt{2\pi} n_{e}^{peak} \sigma_{blob}$, and $\Tau_{\sigma} = \sigma_{blob}/v_0$, where $\sigma_{blob}$ is the spatial standard deviation of the 2D Gaussian distribution describing the electron density in the plasma cloud, and $v_0$ is the cloud's velocity (actual or apparent). The resulting parameters that minimised $\chi^2$ per degree of freedom $\chi^2/\text{ndf}$ (i.e. reduced $\chi^2$) were: $\Delta DM = 0.027 \pm 0.002$ pc cm$^{-3}$, $\Tau_{\sigma} = 12 \pm 1.0$\,days, $t_{peak} = 62.7\pm 0.7$\,days, and $DM_{offset} = 0.005 \pm 0.001$ pc cm$^{-3}$. 

% /home/msok/Desktop/SKA/papers/2024/EDA2_FRBs/20250522_simple_DM_modelling.odt
Based on the fitted Gaussian parameters, physical parameters of the cloud can be estimated. The timescale of the entire DM-increase event is $\Delta$T $\sim 6 \sigma_{blob} \approx$75\,days, and the mean $\bar{\Delta DM}$ can be calculated as:

\begin{equation}
\bar{\Delta DM} = (1 / \Delta T ) \int ( DM(t) - DM_{offset} ) dt \approx0.01\,pc\,cm^{-3}. 
\label{eq_mean_delta_dm}
\end{equation}

% /home/msok/Desktop/SKA/papers/2024/EDA2_FRBs/20250522_simple_DM_modelling.odt
Assuming that the transverse motion of the screen relative to the observer on Earth is dominated by the pulsar's proper motion $v_c$=120\,km/s (see eq.~\ref{eq_velocities} and discussing in Section~\ref{subsec_refractive_scintillation}), we estimate the size of the cloud to be R$= v_c\Delta T \approx 3\cdot 10^{-5}$\,pc, and the resulting mean electron density in the cloud to be $\bar{n_e} \approx \bar{\Delta DM} / R \approx$380 e$^{-}$cm$^{-3}$.

% /home/msok/Desktop/EDA2/papers/2024/EDA2_FRBs20250505_DM_modelling_multiple_cases.odt
% and Figure 3 in 20250423_fitting_DM_and_Tau_variations_PAPER.odt
Based solely on our data, it is not possible to measure $n_{e}^{peak}$, $\sigma_{blob}$, and $v_0$ without additional assumptions. Therefore, we have inferred physical parameters for the case of apparent transverse velocity of the screen $v_0 = 120$\,km/s, and the results are summarised in the middle part of Table~\ref{tab_comparison}. We note that $\sigma_{blob}$ and $n_{e}^{peak}$ for other values of $v_0$ can be calculated by scaling results for $v_0 = 120$\,km/s. For example, $\sigma_{blob}$ and $n_{e}^{peak}$ for $v_0 = 60$\,km/s can be calculated by respectively dividing and multiplying by a factor 120/60.

% OLD TEXT BEFORE MAKING THE PARAMETERS FULLY INDEPENDENT and then discussing cases of 60 and 120 km/s
% where y$= v_{crab} \alpha (t-t_{peak})$ and $\alpha$ is the proportionally constant relating velocity of the ``blob'' with the transverse velocity $v_{crab}$ of the Crab pulsar (120\,km/s \citep{2008ApJ...677.1201K}), and $DM_{offset}$ is a small offset in DM (very close to zero). The parameters resulting from the fit shown in Figure~\ref{fig_dm_model_fit} are: $t_{peak} = 62.7\pm 0.7$\,days, $\sigma_{blob} = 4.5 \pm 1.3 \cdot10^{-6}$\,pc (FWHM$_{blob} = 1.1 \pm 0.3 \cdot 10^{-5}$\,pc) and peak electron density $n_e = 2400 \pm 700$\,$e^{-}/cm^3$. The fitted value of $\alpha$ was $\alpha = 1.07 \pm 0.3$, which corresponds to transverse velocity of the blob $v_b = 130 \pm 40$km/s, which can be interpreted as the velocity of the plasma blob itself or apparent movement of the plasma blob in the vicinity of the Crab PSR due to movement of the pulsar itself. We note that very similar results were obtained when $\alpha$ was fixed as $\alpha=1$.

% /home/msok/Desktop/SKA/papers/2024/EDA2_FRBs/20250522_simple_DM_modelling.odt
The characteristic spatial scale of the plasma blob resulting from the fit is $\sigma_{blob} \approx 4.2 \cdot 10^{-6}$\,pc, and $6\sigma_{blob} \approx 2.5 \cdot 10^{-5}$\,pc is consistent with the cloud size R$\approx 3\cdot 10^{-5}$\,pc estimated above using the mean values. This $\sigma_{blob}$ value leads to peak electron density $n_{e}^{peak} \approx$2600\,e$^{-}$cm$^{-3}$. As an additional check, we performed a standard DM structure function analysis (for description see Section 5.2 in \citet{2018MNRAS.479.4216M}) and obtained a characteristic timescale of 12.9\,days, which corresponds to $4.3\cdot10^{-6}$\,pc (when multiplied by 120\,km/s) and agrees nearly perfectly with the $\sigma_{blob}$ value.

Hence, all tested methods lead to consistent results, and the resulting parameters are physically plausible. Despite the different epochs of observations, frequencies etc., they are also in-line with the studies by \citet{2018MNRAS.479.4216M,2008A&A...483...13K,2011MNRAS.410..499G} as summarised in Table~\ref{tab_comparison}. This suggests that similar scintillation/scattering events due to plasma screens of the size $\sim 10^{-5}$\,pc and electron densities of the order of $10^2$ -- $10^4$\,$e^{-} cm^{-3}$ are relatively common for the Crab pulsar. The observed electron densities are consistent with the filamentary structures inside the Crab Nebula (e.g. \citet{1957PASP...69..227O,1982ApJ...258....1F}).

% TEST TABLE:
% \begin{table*}[hbt!]
% \begin{threeparttable}
% \caption{Comparison of results in this work with earlier studies. The upper part of the table shows quantities related derived from variations in DM, while the lower part of the table shows results derived from the scatter broadening time.}
% \label{table_example}
% \begin{tabular}{|c|c|c|c|c|}
% \toprule
% \multicolumn{2}{c|}
% \headrow \textbf{Quantity} & \multicolumn{2}{c|}{\textbf{This work}}  & \textbf{\citet{2008A&A...483...13K}} & \textbf{\citet{2018MNRAS.479.4216M}}\\
% \midrule
% $\Delta$DM & \multicolumn{2}{c|}{0.03 pc cm$^{-3}$} & 0.03 pc cm$^{-3}$ & 0.10 pc cm$^{-3}$ \\ 
% \midrule
% & 120 km/s & 60 km/s & & \\ 
%\midrule
%\bottomrule
%\end{tabular}
%\end{threeparttable}
%\label{tab_comparison_test}
%\end{table*}

% NEW UPDATE WITH 1400 km/s : /home/msok/Desktop/EDA2/papers/2024/EDA2_FRBs/20250505_DM_modelling_multiple_cases.odt
% 
% NEW/FINAL : Figure 1 in 20250423_fitting_DM_and_Tau_variations_PAPER_FINAL_SIMPLE.odt
% /home/msok/Desktop/EDA2/papers/2024/EDA2_FRBs/20250429_summary_of_size_scales_for_the_paper.odt
% OLD VERSION OF TABLE with 120 and 60 km/s IN : ~/Desktop/SKA/papers/2024/EDA2_FRBs/versions/20250505/0905am_Table_with_120_and_60_kms    
\begin{table*}[hbt!]
\begin{threeparttable}
\caption{Comparison of results in this work (as described in Section~\ref{subsec_modelling_dm_and_tau}) with earlier studies. Our estimates of plasma cloud size and electron densities are very similar to those published by \citet{2008A&A...483...13K},  \citet{2018MNRAS.479.4216M} and \citet{2011MNRAS.410..499G}.}
% The best self-consistency of the estimates on the size of the plasma blob is obtained for the small distance ($\sim$1\,pc) between the pulsar and the blob, which leads to size of the blob $\sim 10^{-5}$\,pc and peak electron density of the Gaussian distributed plasma $\sim 2500 \pm 200$e$^{-}$cm$^{-3}$. 
\label{table_example}
\begin{tabular}{|c|c|c|c|c|}
\toprule
% \multicolumn{2}{c|}
\headrow \textbf{Quantity} & \textbf{This work} & \textbf{\citet{2008A&A...483...13K}} & \textbf{\citet{2018MNRAS.479.4216M}} & 
\textbf{\citet{2011MNRAS.410..499G}} \\
\midrule
$\Delta$DM [pc cm$^{-3}$] & 0.02 & 0.03 & 0.10 & 0.15 \\ 
\midrule
Cloud size based on $\bar{\Delta DM}$ \tnote{a} & $3 \cdot 10^{-5}$ & $7 \cdot 10^{-5}$ & $2 \cdot 10^{-5}$ & $3 \cdot 10^{-5}$ \\
$\bar{n_e}$ [e$^{-}$ cm$^{-3}$] & 380 & $10^3$ -- $10^4$ & Not calculated & $10^4$ \\
\midrule
\multicolumn{2}{c|}{\textbf{Modelling DM(t) variations}} & & & \\
\midrule
 $\sigma_{blob}$ & $4.2 \cdot 10^{-6}$ &  &  & \\
 $6\sigma_{blob}$ & $2.5 \cdot 10^{-5}$ &  &  & \\
 Peak $n_{e}$ [$e^{-}$/cm$^{-3}$] & 2500$\pm$200 & & & \\
% \midrule
% & & & \\
\midrule
\multicolumn{2}{c|}{\textbf{Modelling $\tau$(t) variations}} & & & \\
\midrule
Range of $\tau$ [ms] & 2 -- 5 & 8 -- 27 & 0.1 -- 0.7\tnote{b} & Not measured \\
\midrule
Range of $\tau$ [ms] & 2 -- 5 & 0.8 -- 2.5 & 2.5 -- 15.6\tnote{b} & Not measured \\
scaled to 215\,MHz with $(215/\nu)^{-3.5}$ & \multicolumn{2}{c|}{} & & 
  \\
% \midrule
% & & &  \\
 % \todo{Update to Viss, x, Les:}R based on $\tau$ variations [pc]\tnote{b} & $50 \cdot 10^{-5}$ & $1.5 \cdot 10^{-5}$ (note\tnote{d}) & \\
% OLD : USED values from Figure 1, From Theta2rms L = 1.866 arcsec^2 pc  for 1 and 1000 pc :
% OLD : 2025-05-05 : I've used FWHM calculation there
% & $L_{cs} = 1000$\,pc & $L_{cs} = 1$\,pc & & \\ 
% R based on $\tau$ variations [pc]\tnote{b} & $50 \cdot 10^{-5}$ & $1.6 \cdot 10^{-5}$ & $1.5 \cdot 10^{-5}$ (note\tnote{d}) & \\
% \midrule
\bottomrule
\end{tabular}
\begin{tablenotes}[hang]
\item[]Table notes:
\item[a] Details in Section~\ref{subsec_modelling_dm}.

\item[b] The baseline level of scattering time in this work was around 0.1\,ms which corresponds to 2.5\,ms at 215\,MHz. However, sudden drops down to $\sim$0\,ms were also observed (see Figure 2 in \citet{2018MNRAS.479.4216M}). Since we are not sure if these were valid measurements of extremely low scattering time we used the baseline level as a reference.
% \item[d] This is for 1.5\,pc distance between the plasma blob and the pulsar, they also provided the result $3.2 \cdot 10^{-6}$\,pc for a distance of 0.1\,pc
\end{tablenotes}
\label{tab_comparison}
\end{threeparttable}
\end{table*}

\subsubsection{Effects of refractive scintillation}
\label{subsec_refractive_scintillation}

% Calculations in /home/msok/Desktop/EDA2/papers/2024/EDA2_FRBs/20250505_fitting_and_modelling_Tau_variations_only_NEW-TAU_PAPER.odt
The long observed timescale $\sim$70\,days of the variations in GP rate and fluence distribution is a manifestation of intensity modulations caused by refractive scintillation. Moreover, estimates of the size $\sim 10^{-5}$\,pc of the plasma cloud agree very well with the large-scale structures required by refractive scintillation (see, for example \citet{1999ApJ...514..272B}).
The refractive scintillation effects on Crab GPs and average profiles have been observed and discussed before (e.g. \citet{1995ApJ...453..433L}, \citet{1990MNRAS.244...68R}, and \citet{2011MNRAS.410..499G}). Interestingly, in a recent paper, \citet{2024ApJ...973...87D} discuss the possibility of using GP rate as an ISM probe, which is exactly what transpired from our observations. 

Under this hypothesis and using our equation ~\ref{eq_tau_theta} and equation 4.45 from ~\citet{2012hpa..book.....L} (see also equation 4.1 in \citet{1986MNRAS.220...19R}), we obtain:

% Using the fitted value $\theta_{rms-peak} \approx 0.00965971$\,arcsec, and $L_{ec} = 2000$\,pc, we calculate $V_{iss} = 1300 \pm 100$\,km/s.
\begin{equation}
V_{ISS} = \frac{\theta_{rms} L_{es}}{t_{RISS}},
\label{eq_viss}
\end{equation}

where $t_{RISS}$ is the timescale of the refractive scintillation,  $L_{es}$ is the distance from the Earth to the screen (Fig.~\ref{fig_plasma_blob_model}), $\theta_{rms}$ is the standard deviation of the scattering angle, and $V_{ISS}$ can be interpreted as the velocity of the refractive scintillation pattern at the location of the observer (i.e. on Earth). 

Additionally, the scatter broadening time is related to $\theta_{rms}$ by the equation (e.g. \citep{2012puas.book.....L}) : % Eq 21.21 in Section 21.9 

% No need for a coin : see J-P's and Kevin Koay paper where the 2 becomes clear due to D*D/(D+D) optimising scattering -> D/2 distance of the screen (middle way through).
% \todo{WARNING : in Pulsar Astronomy by A. Lyne it's 1/(4c) vs. 1/(2c) in Kuzmin, vs. in Pulsar Handbook it's just $\theta^2$ d/c !!!??? So, factor (1/2) or (1/4) difference !!! Who can be trusted - use a coin ???}
\begin{equation}
\tau = \frac{L_{es} \theta_{rms}^2}{2c},
\label{eq_tau_theta}
\end{equation}

where c is the speed of light. 

% /home/msok/Desktop/EDA2/papers/2024/EDA2_FRBs/20250524_proper_estimates_of_deltaTriss.odt
Typically, scintillation analysis (e.g. \citet{2014ApJ...791L..32B}), uses auto-correlation function (ACF) of dynamic spectrum containing scintles to calculate scintillation time and de-correlation bandwidth as respectively the full-width at 1/e and half-maximum half-width of the 2D Gaussian fitted to the correlation peak in ACF. In this analysis the fluence normalisation or number of GPs were sampled daily in 1\,hour observations. Hence, long-timescale dynamic spectrum is not available, and we can only use 1D temporal dependence of the number of GPs or normalisation of the fluence distribution ($N_{ref}$). Hence, in order to be consistent with this convention, we estimate $t_{RISS}$ by fitting Gaussian to the peak of the normalisation $N_{ref}$ of the fluence distribution (not shown here but very similar to Figure~\ref{fig_ngp_vs_date}) as a function of time, and calculate $t_{RISS} \approx$ 51\,days as full width at 1/e of the peak. 

In order to calculate $V_{ISS}$ using equation~\ref{eq_viss}, the distance to the screen $L_{es}$ and $\theta_{rms}$ are also required. The additional constraints are imposed by equation~\ref{eq_tau_theta} (above) and the following equation (equation 3 in \citet{2014ApJ...791L..32B}):

\begin{equation}
V_{ISS} = |x v_{c,\perp} + v_{e,\perp} - (1+x)v_{s,\perp}| ,
\label{eq_velocities}
\end{equation}

% NEW : /home/msok/Desktop/EDA2/papers/2024/EDA2_FRBs/20250522_simple_Tau_modelling_FINAL.odt
% /home/msok/Desktop/EDA2/papers/2024/EDA2_FRBs/
% 20250522_simple_Tau_modelling.odt
% 20250521_new_discussion_viss_x_etc.odt
where x=$L_{es}/L_{sc}$ defines the location of the screen, v$_{c} \approx$120\,km/s is the pulsar's proper motion, v$_{e} \approx$ 30\,km/s is the Earth's proper motion and v$_{s}$ is the proper motion of the screen. We note that from now onwards we shall use the known velocities of the pulsar and Earth as the case of maximum transverse velocity, and consequently the $\perp$ symbol will be dropped. The combination of these three equations leads to the following 3-order equation in y$:=\sqrt{L_{es}}$:
% Typical screen velocities are $v_s \sim$15\,km/s \citep{2014ApJ...791L..32B,1994A&A...287..390B} are much lower than pulsar proper motion, and we neglect this term in further analysis.

\begin{equation}
\alpha \text{y}^3 + (\Delta_1 - \Delta_2) \text{y}^2 - \alpha L_{ec}\text{y} -L_{ec} \Delta_2 = 0,
\label{eq_third_order}
\end{equation}

where $\alpha = 1574.85\sqrt{\tau_{ms}}/t_{RISS}^{days}$, $\Delta_1 = (v_c - v_{s})$, and $\Delta_1 = (v_e - v_{s})$.
For the maximum scattering time $\tau=$5\,ms and $t_{RISS}=$51\,days, this equation has 2 physically acceptable solutions, $y_1 = 0.22$, and $y_2 = 42.3$, which lead to $L_{es,1} =$0.05\,pc (screen close to Earth), and $L_{es,2} =$1835\,pc (screen close to the pulsar). 

The first solution $L_{es,1} =$0.05\,pc could potentially be related to the inner edge of the solar system (distance $\sim$0.01\,pc), but this does not have convincing physical explanation. On the other hand, the second solution ($L_{es,2} =$1835\,pc) places the screen close to the pulsar, and the distance matches the distance to the Perseus spiral arm of the Milky Way galaxy, where the Crab pulsar is located. Hence, the scattering screen could be placed somewhere at the edge of the Perseus arm as seen along the LoS from Earth, which was also proposed by earlier studies (e.g. \citet{1971ApJ...166..513C}). Alternatively, given the accuracy of this simplistic approach, and uncertainties of the distance to the pulsar, velocities of the screen and pulsar it is possible that the scattering material corresponds to the expanding Crab Nebula surrounding the pulsar. We note that solar elongations of Crab pulsar during the observing period were between 180\degree and 74\degree - sufficiently large to make any contributions from the solar wind negligible.
% /home/msok/Desktop/EDA2/papers/2024/EDA2_FRBs/20250527_solar_elongation.odt

% /home/msok/Desktop/SKA/papers/2024/EDA2_FRBs/20250522_simple_Tau_modelling_FINAL.odt
\begin{table}[hbt!]
\begin{threeparttable}
\caption{Parameters $\theta_{rms}$, $L_{es}$, x, and $V_{ISS}$ calculated using equation~\ref{eq_third_order} applied to scattering times $\tau=$2,3,4 and 5\,ms as observed in our data.}
\label{tab_results_for_different_taus}
\begin{tabular}{ccccc}
\toprule
\headrow $\tau$[ms] &  $\theta_{rms}$[arcsec] & $L_{es}$[pc] & x & $V_{iss}$[km s$^{-1}$] \\
\midrule
% NEW : distance = 1.9kpc :
% /home/msok/Desktop/EDA2/papers/2024/EDA2_FRBs/20250522_simple_Tau_modelling_FINAL_D1p9kpc.odt
2 & 0.030 & 1797 & 17.5 & 1851 \\ 
3 & 0.037 & 1816 & 21.6 & 2279 \\ 
4 & 0.040 & 1827 & 25.0 & 2640 \\ 
5 & 0.050 & 1835 &  28.0 & 2957 \\ 
% Distance = 2kpc :
% /home/msok/Desktop/EDA2/papers/2024/EDA2_FRBs/20250522_simple_Tau_modelling_FINAL.odt
% 2 & 0.030 & 1894 & 18.0 & 1900 \\ 
% 3 & 0.036 & 1916 & 22.1 & 2340 \\ 
% 4 & 0.040 & 1925 & 25.6 & 2710 \\ 
% 5 & 0.050 & 1933 &  28.8 & 3035 \\ 
\bottomrule
\end{tabular}
% \begin{tablenotes}[hang]
% \item[]Table note
% \item[a]First note
% \item[b]Another table note
% \end{tablenotes}
\end{threeparttable}
\end{table}

The above estimates were obtained for the specific maximum scattering time $\tau=$5\,ms. Hence, the question arises as to what screen distance can be calculated for other scatter broadening times (i.e. different dates). We calculated $L_{es}$ for other values to $\tau_{ms}$ between 5 and 2\,ms (Tab.~\ref{tab_results_for_different_taus}). The distance to the screen varies from about 1835 down to 1797\,pc. Obviously, if increase in scatter broadening time is caused by a single plasma cloud, the distance to this cloud should remain constant within the measurements errors. However, we believe that the different distance values in Table~\ref{tab_results_for_different_taus} are due to measurement errors and very simplistic modelling approach. For example, orientation of all velocity vectors has been ignored, and only maximum velocity values have been used in equation~\ref{eq_third_order}. Simple tests show that even small changes in $v_e$ or $v_s$ corresponding to changing orientation to the Earth's proper the Sun motion are of the similar order to the observed differences (Tab.~\ref{tab_results_for_different_taus}) in $L_{es}$, $V_{ISS}$ etc. A more complex modelling calculating $L_{es}$ for various orientations of velocity vector, screen velocities, and propagation of errors to solutions of equation~\ref{eq_third_order} should be performed to further validate these statements and results. We defer more detailed modelling to future work.

% \todo{revise/include error analysis - variations of direction etc ???:}
% values are within the errors due to errors on $\tau$, $L_{ec}$ and velocities of the pulsar and Earth. Therefore, we conclude that this is a plausible physical solution. We note that with a self-consistence model the equation~\ref{eq_third_order} should provide the same distance to the screen for different scattering times. We believe, that the unaccounted changes in orientation may lead to the differences, and once the actual perpendicular velocities are used for specific days the distance may become closer to constant as even a small change of $v_{c}$ of 20\,km/s in equation~\ref{eq_third_order} changes distance to the screen by 20\,pc. Clearly more accurate modelling is required, which we defer to potential future work.

\subsubsection{Modelling scatter broadening time variations}
\label{subsec_modelling_scatter_broadening_time}

We have also made an attempt to model time dependence of $\tau(t)$ as the following Gaussian:

\begin{equation}
\tau(t) = \tau_0 + \frac{L_{es} \theta_{rms-peak}^2}{2c} e^{-(t-t_{peak})^2/2T_\sigma},
\label{eq_tau}
\end{equation}

where $\tau_0$ is offset, and $T_\sigma$ is the timescale of the variations.
The free parameters of the fit were $\tau_0$, $T_\sigma$ and $L_{es} \theta_{rms-peak}^2$. We note that $L_{es}$ is unknown, and therefore the product $L_{es} \theta_{rms-peak}^2$ was used as a free parameter of the fit. 

\begin{figure}[hbt!]
\centering
% \includegraphics[width=0.98\linewidth]{Tau_vs_DATE_20250502_SNR10_GIMPED.png}
% /home/msok/Desktop/EDA2/papers/2024/EDA2_FRBs/
% NEW : 2025-05-26 : 20250505_fitting_and_modelling_Tau_variations_only_NEW-TAU_PAPER_FINAL2.odt
% OLD : 2025-05-24 : Constant T_sigma=12.46 : 20250505_fitting_and_modelling_Tau_variations_only_NEW-TAU_PAPER_FINAL.odt
% OLD : 2025-05-04 - only theta^2 fitted (not Theta^2 * L ):
% OLD : Figure 1 in 20250505_fitting_and_modelling_Tau_variations_only_NEW-TAU_PAPER.odt
% /home/msok/Desktop/SKA/papers/2024/EDA2_FRBs/PAPER/DM_vs_TIME/20250502/images/FINAL/Tau_vs_DATE_20250502_SNR10_FitDknown.png
% OLD : Figure 2 on page 3 of /home/msok/Desktop/SKA/papers/2024/EDA2_FRBs/20250502_fitting_and_modelling_Tau_variations_only_NEW-TAU.odt
% 
\includegraphics[width=0.98\linewidth]{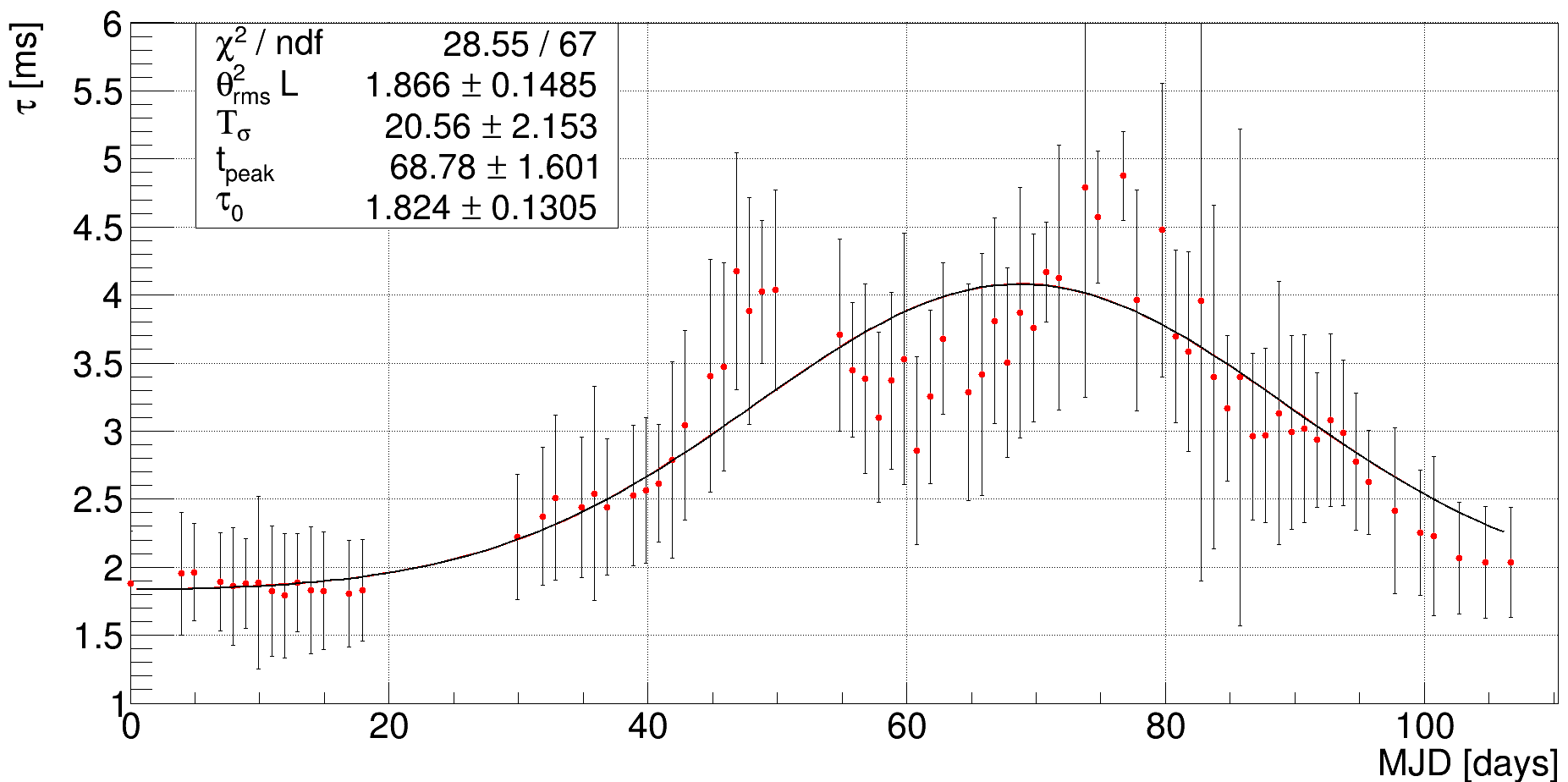}
\caption{Scatter broadening ($\tau$) vs. time fitted with equation~\ref{eq_tau} describing a plasma sphere passing transversely through the LoS between the pulsar and Earth. All data points were used for this fit. The two peaks (``horns'') at around 50 and 75\,days from the start are likely due to smaller scale plasma structures within the screen, and were not full investigated and modelled here.}
% The function was fitted to the overall trend excluding these two ``sub-events''.
\label{fig_tau_model_fit}
\end{figure}

This model can be interpreted as the LoS passing through different parts of the plasma cloud with different electron densities and amount of turbulence causing variable scattering (represented by $\theta_{rms}$). The fitted curve and parameters are shown in Figure~\ref{fig_tau_model_fit}. Although $t_{peak}$ is consistent with the earlier DM modelling, these results show that the timescale $T_\sigma \approx20.6$\,days of $\tau$ variations is slightly longer than fitted to DM variations ($\approx$12.5\,days). Fitting the data with the $T_\sigma$ parameter fixed to 12.5\,days (as obtained from fit to DM(t)) resulted in much worse reduced $\chi^2$ value. 

Hence, the timescale of scattering the scattering is longer than the timescale of DM variations. Also, the scattering started and finished before the contribution of the plasma cloud to the total DM became measurable. It is expected that the onset of scattering becomes observable before the direct LoS is crossed by the cloud, and similar effect was observed by \citet{2008A&A...483...13K}.   

We interpret our findings (Sections~\ref{subsec_refractive_scintillation} and ~\ref{subsec_modelling_scatter_broadening_time}) using two screens model. With the variable component of the scattering caused by a screen in the expanding Crab Nebula or in the Perseus arm of the Galaxy, while the approximately constant ``baseline'' ($\tau \approx$2\,ms) due to the second screen (not modelled here). Such a two screen model was suggested and applied by the earlier studies \citep{1971ApJ...166..513C,1977ApJ...214..214I,1975MNRAS.172...97L,1976ApJ...209..578V,1993ApJ...410..673G,2021ApJ...915...65M}. The peaks (``horns'') in scattering time at around 43 -- 58 and 65 -- 83 days, may be due to additional sub-structures (lumps) of turbulent plasma within the main cloud. The fit to the data excluding these structures yielded very similar results. However, we defer more complex modelling using the two screen model, and potential inclusion of the ``horns'' to the future work.

\subsubsection{Spectral behaviour of scatter broadening}
\label{subsec_tau_freq_scaling}

\begin{figure*}[hbt!]
\centering
\includegraphics[width=0.45\linewidth]{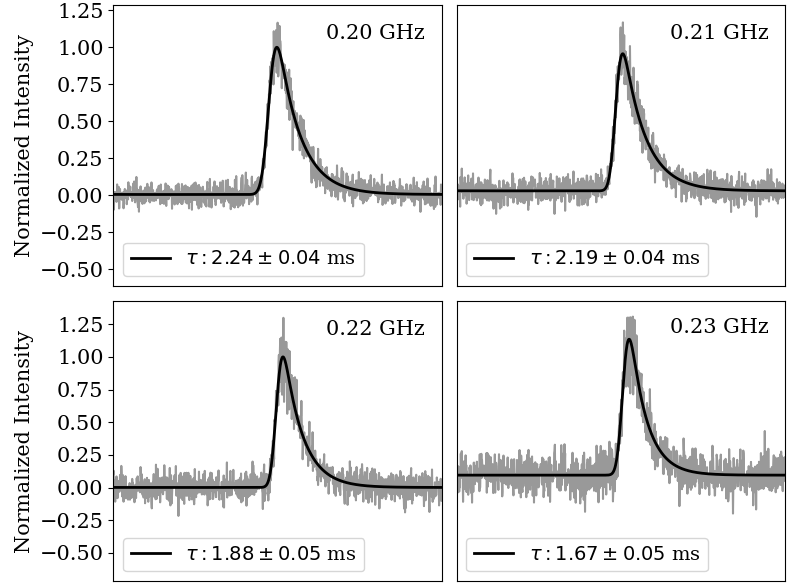}
\includegraphics[width=0.45\linewidth]{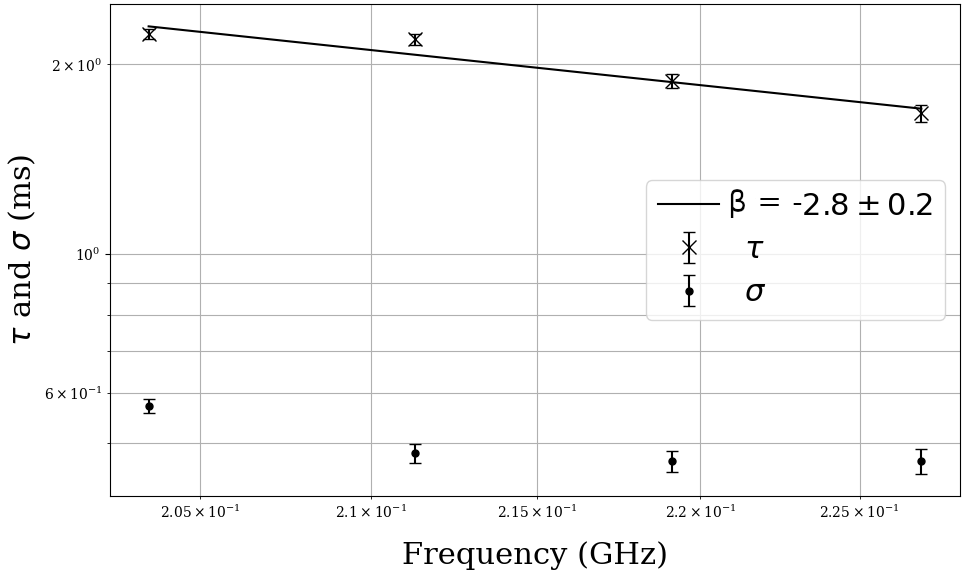}
\caption{Results of scattering modelling using SCAMP package (Sec.~\ref{subsec_tau_freq_scaling}). Left panel: pulse profiles fitted in four sub-bands to maximum SNR GP recorded on 2024-12-24. Right: scattering time $\tau$ resulting from these fits as a function of frequency with the value of spectral index $\beta = 2.8\pm0.2$. $\sigma$ is the standard deviation of the Gaussian profile representing intrinsic (unaffected by scattering) pulse width. The figures and units (frequency in GHz) are exactly as generated by the SCAMP package.}
\label{fig_scamp_results}
\end{figure*}

\begin{figure}[hbt!]
\centering
\includegraphics[width=0.95\linewidth]{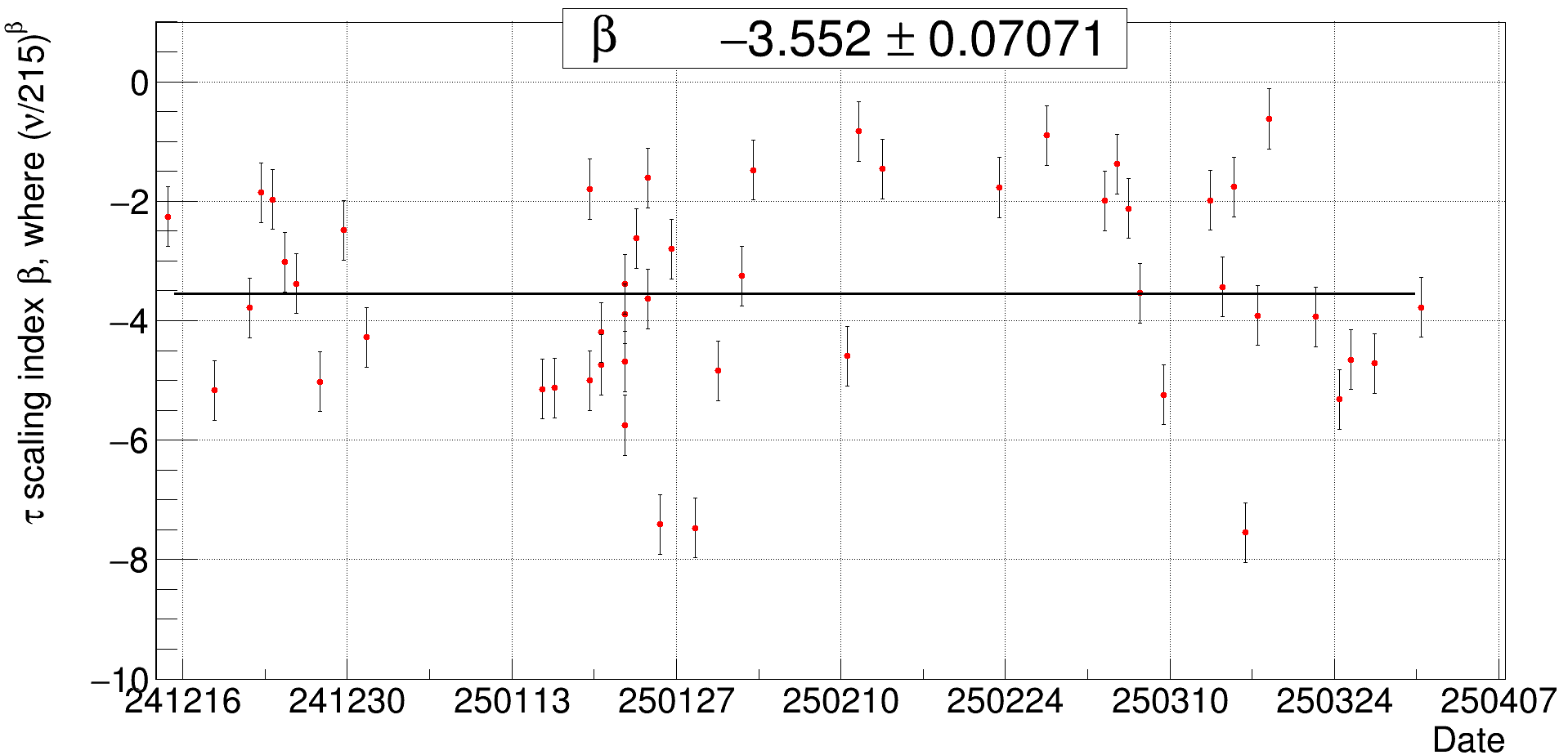}
\caption{Spectral index of the scatter broadening time as a function of time after excluding outliers due to bad data or fit convergence ($\beta>0$ and $\chi^2/\text{ndf}>3$).}
\label{fig_beta_vs_date}
\end{figure}

We have also made an initial attempt to measure frequency scaling of scatter broadening time using bright GPs and 
software package SCAMP \footnote{\url{https://github.com/pulsarise/SCAMP-I}} \citep{2021MNRAS.504.1115O,2016MNRAS.462.2587G}. 
The band was divided into 4 sub-bands and MCMC fitting of individual GP profiles was performed in each sub-band (see example in left Figure~\ref{fig_scamp_results} for the brightest GP from 2024-12-24), and yielded scatter broadening times ($\tau$) consistent with our method. Based on this, scattering spectral index $\beta$ (where $\tau \propto (215/\nu)^{\beta}$) was fitted to scattering time $\tau$ as a function of frequency (e.g. right image in Figure~\ref{fig_scamp_results}).
Figure~\ref{fig_scamp_results} shows that the package performs well on Crab GPs despite the fact that it was originally designed to work on average pulse profiles.
The same fitting procedure was applied to all the datasets and yielded $\beta$ as a function of time. The fitting performs very well during low scattering conditions for high SNR GPs, but did not converge well when the scattering was higher (SNR lower). Several outliers and failed fits were identified and discarded, and spectral index of the scatter time as a function of date is shown in Figure~\ref{fig_beta_vs_date}. The mean value of the spectral index is $-3.6 \pm 0.1$, which is in a good agreement with the previous measurements with the MWA (e.g. \citet{2019ApJ...874..179K}), \old{and broadband measurements by \citet{2006ARep...50..562P,2002AstL...28..251K}.}
Overall the spectral index varies significantly on various dates. Especially, the SCAMP fit did not perform well during high-scattering conditions regardless if isotropic or anisotropic scattering model 
was used. This could be caused by either the reduced SNR or neither of the models being applicable to these high scattering conditions. Additionally, the flattening of scattering spectral index observed when scattering was high (mid-February and early March 2025) may be a result of inadequate screen model.

These results show that any measurements of scattering spectral index $\beta$ for Crab pulsar have to be treated with caution.
Mainly because highly variable scattering conditions may impact the fitting results on a particular day (e.g. cause apparent flattening) by reducing SNR or requiring different screen models (e.g. thin vs. thick, isotropic vs. anisotropic etc.). Additionally, as noted in \citet{2021MNRAS.504.1115O,2016MNRAS.462.2587G}, the spectral index $\beta$ measurements are strongly dependent on the fitting procedure or package (e.g. fitting in log-log space vs. original values etc.).
It is clear that consistent measurements require further work on screen models and more stable fitting procedures, especially during high-scattering/low-SNR conditions. We leave these efforts for the future, and we are willing to contribute to adjustments in the SCAMP package to make it suitable for single pulse analysis.

% ~/Desktop/EDA2/papers/2024/EDA2_FRBs/PAPER/SCATTERING_SPECTRAL_INDEX

% Figure 9 in 20250529_Tau_using_SCAMP-I_package_ALL_DATA_TauVsTime_FINAL.odt
% /media/msok/f5ce6064-9dc3-40c3-bc9c-7314e8594519/eda2/images/FINAL/Tau_vs_date_PAPER.png 
%  ~/Desktop/EDA2/papers/2024/EDA2_FRBs/PAPER/SCATTERING_SPECTRAL_INDEX/Tau_vs_date_PAPER.png
% 

\subsection{FRB context}
\label{subsec_frb_context}

% /home/msok/Desktop/SKA/papers/2024/EDA2_FRBs/20250426_number_of_FRBs_from_CRAB_GPs.odt
The possibility of extra-galactic young Crab-like pulsars being responsible for FRBs has been discussed by multiple authors (see references in Section~\ref{sec_introduction}). Although, the energies, distribution of times between pulses, and other arguments favour magnetars as the progenitors of majority of FRBs, in this section we revisit this possibility using our huge sample of low-frequency GPs.  

\subsubsection{Spectral luminosity}
\label{subsec_spectral_luminosities}

The maximum spectral luminosities observed in our sample are of the order of $5 \cdot 10^{25}$ ergs/s/Hz (right plot in Fig.~\ref{fig_flux_density_distribution}), which is consistent with earlier measurements performed typically at higher frequencies (e.g. Figure 3 in ~\citet{2022NatAs...6..393N}). These spectral luminosities  are some 3 orders of magnitude lower than the faintest pulses from repeating FRB 20200120E reported therein. By extrapolating the fitted spectral distribution of Crab GPs (right plot in Figure~\ref{fig_flux_density_distribution}) to higher luminosities, we find that the waiting time for $\sim 10^{28}$erg/s/Hz pulse is of the order of $T_{28} \sim$\,66,000 years. The total number of galaxies up to redshift z$\le$8 was estimated by \citet{2016ApJ...830...83C} to be $N(z<8) \sim10^{12}$. We estimate number galaxies up to redshift z=0.1 to be $N(z<0.1) \sim 10^8$ (the highest reported FRB redshifts are $\sim$1 \citep{2023Sci...382..294R}) by scaling $N(z<8)$ by the ratio of co-moving volumes \footnote{\url{https://www.astro.ucla.edu/~wright/CosmoCalc.html}}, while $N(z<1) \sim 10^{10}$. We note that, scaling of $N(z<8)$ by the ratio of co-moving volumes is a simplification not taking into account redshift evolution of the number of galaxies. Next, we assume number of young Crab-like pulsars per galaxy $N_{\text{yp}}\sim 1$. There are no good estimates for this number and it could be as low as a $\sim$1 or as high as $\gtrsim$100. Then a nearby FRB (z<0.1) with spectral luminosity $\sim 10^{28}$erg/Hz/s could be observed every few hours. Hence, FRBs similar to repeating FRB 20200120E, for which Crab-like pulsars are also considered as a potential progenitors \citet{2021ApJ...910L..18B}, could be relatively common. In fact, limits on number of pulses from such objects could provide estimates on the number of Crab-like pulsars. 
% see details in /home/msok/Desktop/EDA2/papers/2024/EDA2_FRBs/20250531_number_of_galaxies.odt

% be in the order of  $1000$\,bln years which is 
Nevertheless, spectral luminosities of Crab GPs are very difficult to reconcile with some 15 orders higher spectral luminosities of typical one-off FRBs. Even assuming that the energy ``reservoir'' in Crab pulsar is sufficient to produce pulses $\gtrsim 10^{34}$ ergs/s/Hz, the expected FRB wait times would be about several orders of magnitude longer than the Hubble time. Hence, our observations are in agreement with the current models that Crab-like pulsars may be responsible for a population of weaker repeating FRBs, but cannot produce more energetic FRBs. 

% Lambda vs. Time : /home/msok/Desktop/EDA2/papers/2024/EDA2_FRBs/20250428_distrib_time_interval_between_pulses_vs_TIME_PAPER-NEW.odt

\subsubsection{Arrival times}
\label{subsec_arrival_times}

Several studies \citep{2024ApJ...973...87D,1995ApJ...453..433L,2010A&A...515A..36K} showed that the distribution of pulse separation times for the Crab pulsar follows an exponential distribution, which corresponds to a purely random, Poisson, process. Hence, the detection of a pulse does not have impact on waiting time for the next one. This differs from what is observed in repeating FRBs, in which case the distribution is better characterised by Weibull distribution resulting from some form of correlation (clustering or anti-clustering) between the subsequent pulses (e.g. in clustering, detection of a pulse increases the probability of detecting another pulse). 
The distribution of time separations of GPs was studied before  \citep{1995ApJ...453..433L,2010A&A...515A..36K,2024ApJ...973...87D}. However, these earlier 
studies were performed at frequencies above 800\,MHz (mainly L-band), and in some cases (e.g. ~\citet{1995ApJ...453..433L}) only up to a time difference of about 66\,s (2000 pulsar periods). \red{Additionally, recent high-frequency (1.6\,GHz) observations by \citet{2023ApJ...959..111L} reported detections of multiple (up to 6) micro-bursts within a single pulse window ($\lesssim$1\,ms), largely exceeding numbers expected from a purely independent process. At our frequencies, however, micro-bursts within a pulse window (or even a few ms) are ``blended'' together due to large scatter broadening time. Hence, we used our giant sample of GPs to verify the distribution of pulse time separations at time scales longer than the pulsar period (verification of our merging procedure confirmed lack of pulses separated by less than 50\,ms).}

\begin{figure*}[hbt!]
\centering
% Final after using only N_gp>2000 datasets, and longer than 3000 seconds:
% /home/msok/Desktop/EDA2/papers/2024/EDA2_FRBs/20250910_distrib_time_interval_between_pulses_PAPER-REVIEW2-PLOTS.odt
% 
\includegraphics[width=0.45\linewidth]{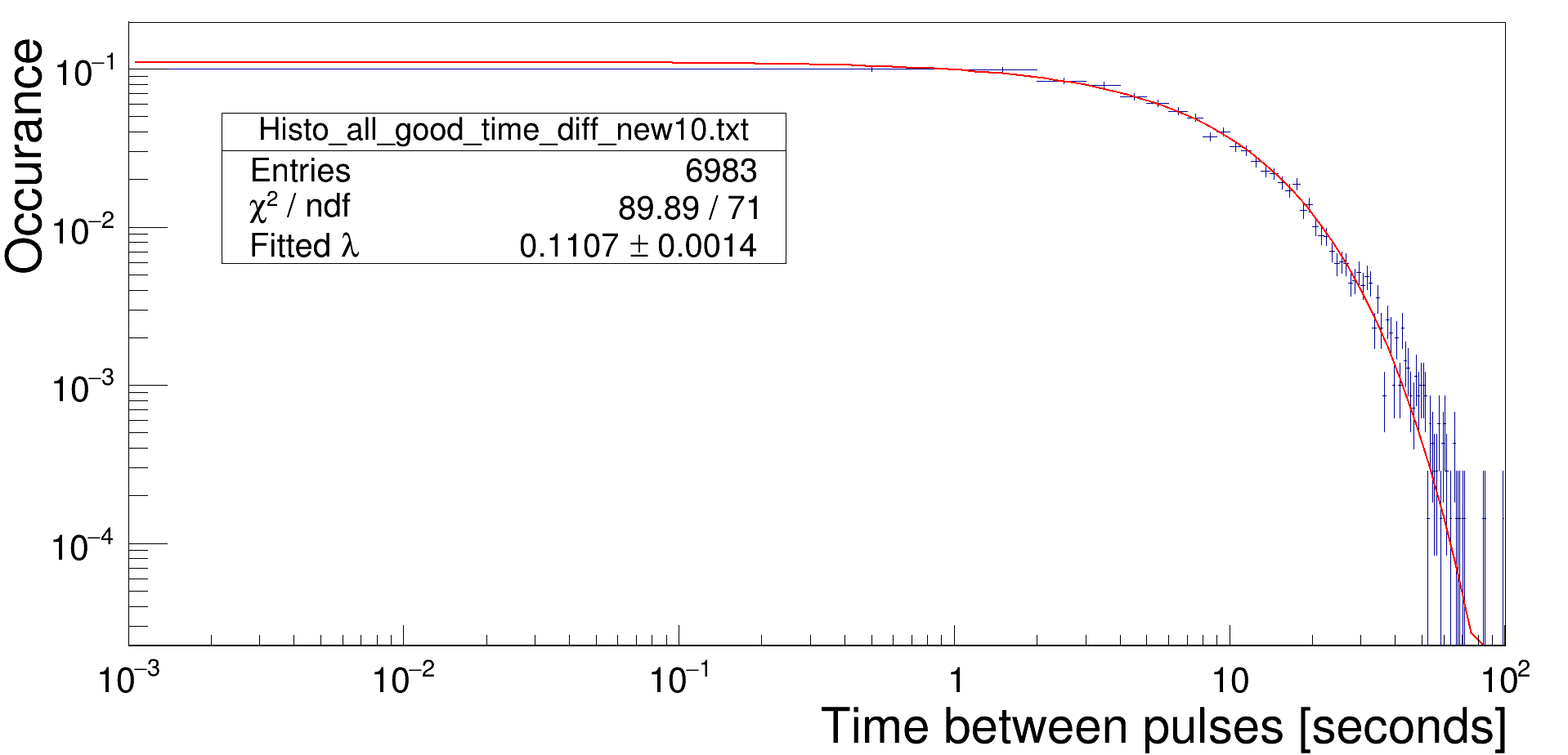}
\includegraphics[width=0.45\linewidth]{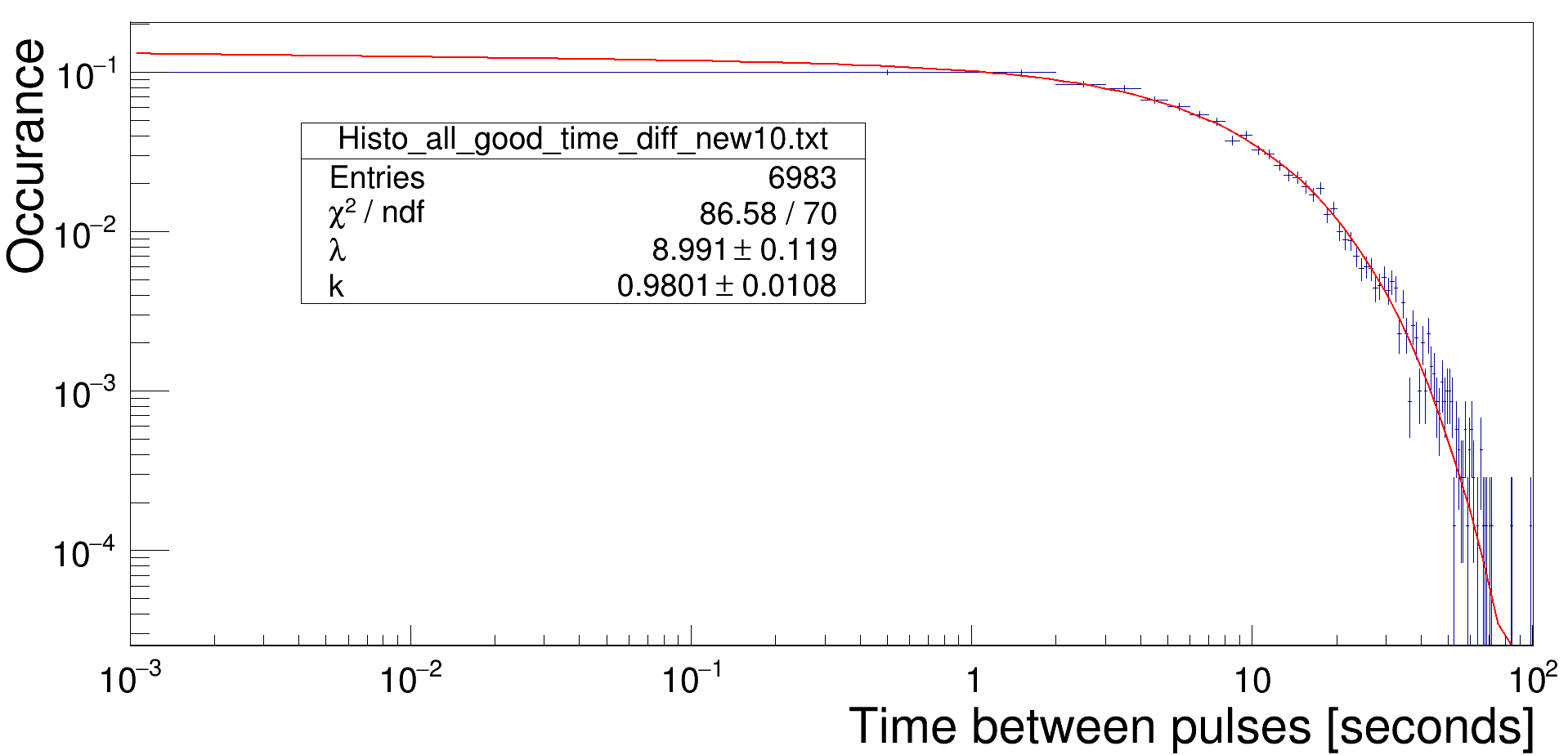}
\caption{\red{Distribution of time between GPs with SNR$\ge$10 where the full sample was used, but the time separation was only calculated within the 1-hour datasets (recorded daily). Only ``good datasets'', with at least 2000 GPs and at least 3000\,seconds of non-flagged data, were used in order to avoid biases caused by high scatter broadening time (i.e. very small number of GPs detected). Left: fitted with exponential distribution (eq.~\ref{eq_exponential_distrib}). Right: fitted with Weibull distribution (eq.~\ref{eq_weilbul_distrib}). Both fits are consistent with the exponential distribution, and there is no indication of clustering or anti-clustering of GPs on timescales $\gtrsim$50\,ms.}}
% \todo{No need to separate MPs and IPs this will make little difference as they are max 0.033/2.00 apart I do not care about such accuracy. Try to add my simulation here !}
\label{fig_time_diff_distribution}
\end{figure*}

Figure~\ref{fig_time_diff_distribution} shows distributions of time between pulses for GPs with SNR$\ge$10, \red{where we only used ``good data sets'' (at least 3000\,s of non-flagged data with at least 2000 GPs detected) to probe time differences up to 600\,s calculated within each dataset only (up to 1\,h observations). This ensured that the gaps between the observations or refractive scintillation did affect the analysis.} The SNR$\ge$10 threshold ensures that false-positives candidates, which can significantly impact the distribution, were not included. \red{However, we verified that the SNR thresholds higher than 7 give consistent results. The resulting distribution can be very well-fitted with an exponential distribution:}

\begin{equation}
e(t) = 
\begin{cases}
  = \lambda e^{-\lambda t} \text{ for t }\ge 0 \\
  = 0 \text{ for t } < 0, \\
\end{cases}
\label{eq_exponential_distrib}
\end{equation}

resulting in \red{$\lambda = 0.111 \pm 0.001$} (left image in Figure~\ref{fig_time_diff_distribution}). The same fit was performed for each separate dataset, and in all cases the distribution of time between pulses was well described by an exponential distribution. \red{The value of the fitted parameter $\lambda$ varies between $\approx0.15$ when the number of GPs detected per hour was highest (end of December 2024) down to $\approx$0.09 when the GP rate was lower (end of March 2025). }The plot of fitted $\lambda$ vs. time has similar shape to Figure~\ref{fig_ngp_vs_date}, and has not been included here for brevity. This varying behaviour of $\lambda$ is perfectly in line with the fact that $\lambda$ represents number of GPs in a certain time interval (GP rate) which was changing with changing scattering conditions (as discussed in Section~\ref{subsec_num_gps_vs_time}). 

\red{The distribution was also fitted with a Weibull distribution (right image in Figure~\ref{fig_time_diff_distribution}) :}

\begin{equation}
w(t) = 
\begin{cases}
  = \frac{k}{\lambda} \left(\frac{t}{\lambda}\right)^{k-1} e^{-(t/ \lambda)^k} \text{ for t }\ge 0 \\
  = 0 \text{ for t } < 0, \\
\end{cases}
\label{eq_weilbul_distrib}
\end{equation}

which reduces to the exponential distribution for shape parameter k=1, and $k<1$ corresponds to clustering. \red{The fit yielded k = $0.98 \pm 0.01$ consistent with 1, which confirms that the distribution of pulse arrival times is well fitted with exponential distribution as expected for independent events (i.e. Poisson process). }

% \todo{Can I do some other statistical test or MC similation to show that this is Poisson distribution. See this Crab paper that I reviewed at PFT meeting. What about right plot SNR$\ge$10 ???}

% ~/Desktop/SKA/papers/2024/EDA2_FRBs/references/Shradha

\begin{figure}[hbt!]
\centering
\includegraphics[width=0.98\linewidth]{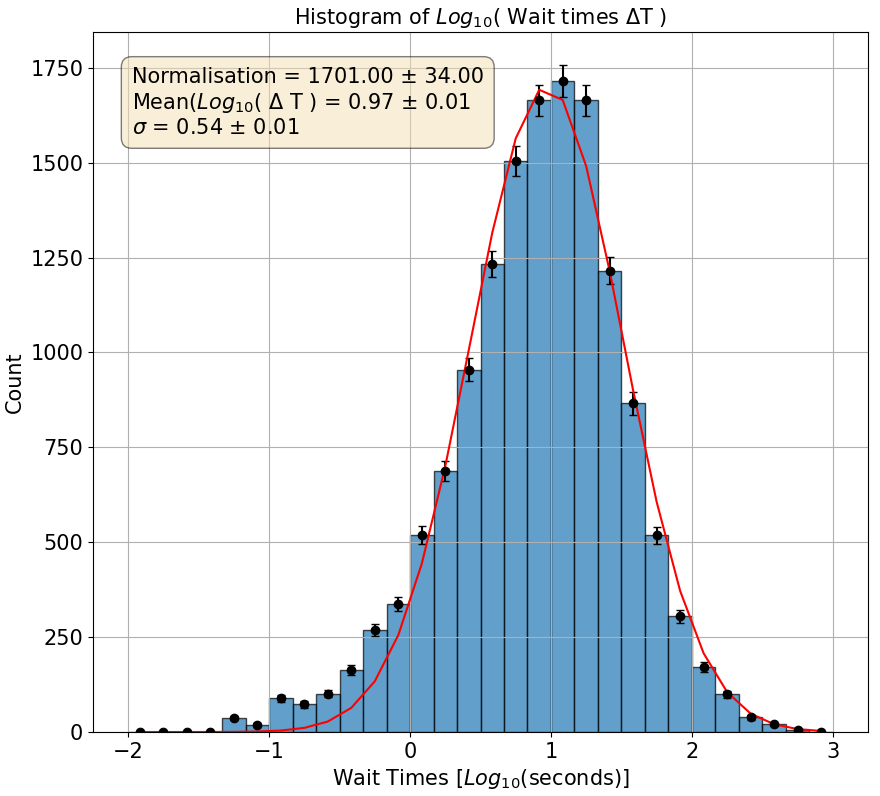}
\caption{\old{Distribution of logarithm to the base 10 of the measured wait times. The observed standard deviation ($\sigma$) of the log-normal distribution fitted to the data is $\sigma_{crab} = 0.54 \pm 0.01$, which fits in between 0 expected for a purely periodic and 0.723 expected for a shot noise process. However, it is lower than $\sigma_{frb} \sim 1.2$ observed in repeating FRBs with sufficiently large number of pulses detected, and much lower than even $\sigma_{sgr}\sim3$ observed in galactic magneters (see Table 1 in \citet{2024OJAp....7E..43K}). This further confirms that statistical properties of arrival times of Crab GPs are different to those observed in repeating FRBs indicating different emission mechanisms and physical processes.}}
\label{fit_log_normal_wait_times}
\end{figure}

% NEW : /home/msok/Desktop/EDA2/papers/2024/EDA2_FRBs/20250612_distrib_time_interval_between_pulses_LOG-NORMAL_DISTR_PAPER.odt
% ~/Desktop/SKA/papers/2024/EDA2_FRBs/PAPER/POISSON_vs_WEIBULL/LOG_NORMAL
% ~/Desktop/SKA/papers/2024/EDA2_FRBs/submission/REVIEW/LOG_NORMAL
% ~/Desktop/SKA/papers/2024/EDA2_FRBs/submission/PostMortem/Jonathan_Katz/
% Two FRBs in the paper 2303.05578v3.pdf are both are much brigheter than Crab:
% FRB 20201124A spectral luminosity ~ 10^34 erg/s/Hz
% FRB 20121102 spectral luminosity ~  from 10^30 to 10^34 erg/Hz/s, with peak luminosities reaching 10^31 erg/Hz/s
\blue{Finally, we have also created a distribution of logarithm to the base 10 of the wait times (Fig.~\ref{fit_log_normal_wait_times}), and compared the standard deviation of the fitted log-normal distribution to those expected for a periodic process, shot noise and observed in repeating FRBs (see Table 1 in \citet{2024OJAp....7E..43K}). The standard deviation of the log-normal distribution fitted to our data $\sigma_{crab} = 0.54 \pm 0.01$ is between 0 expected for a purely periodic process (e.g. regular pulses from a pulsar) and 0.723 expected for shot noise. However, it is lower than $\sigma_{frb} \sim 1.2$ observed in repeating FRBs with sufficiently large number of pulses detected, and much lower than even $\sigma_{sgr}\sim3$ observed in galactic magneters (see Table 1 in \citet{2024OJAp....7E..43K}). This further confirms that statistical properties of arrival times of Crab GPs are different to those observed in repeating FRBs, which also indicates differences in physical emission mechanisms and processes.}

\subsubsection{Limits on low-frequency FRB rate}
\label{subsec_frb_rates_limits}

The entire presented project was originally motivated by searches for pulses from repeating FRBs, magnetars and RRATs. Hence, during this project three repeating FRBs were observed namely:  FRB20240114A (27\,hours), FRB20201130A (27\,hours), FRB20180301 (9\,hours) and FRB20240619D (5\,hours). No pulses at DM of these FRBs were detected above the detection threshold of $\approx$240\,Jy\,ms.

Similarly, no FRB-like events other than Crab giant pulses were detected in the entire dataset (repeating FRBs, RRATs, magnetars, Crab pulsar, nearby galaxy NGC 3109, and several other fields) totalling to about 100\,hours of observations with the station beam FoV$\approx$0.012\,srad. This negative result can be translated into an upper limit on FRB rate R$_{frb}$ in 200 -- 231.25\,MHz band, which is R$_{frb}\le$2500 FRBs/day/sky. This limit is some 2 orders of magnitude higher than the current extrapolations of FRB rates from higher frequencies (Figure 3 in \citet{2024PASA...41...11S}). Hence, the observing time and/or FoV need to increase substantially in order to routinely detect low-frequency FRBs. For example, even increasing number of beams to 10 could lead to first FRB detections after some 1000 hours of observations (about 6 weeks of non-stop observing with EDA2). However, millisecond all-sky imaging, as proposed by \citet{2022aapr.confE...1S}, is the best way to convert EDA2 and SKA-Low stations into ``FRB machines'' once the technical challenges of data capture, high-time resolution imaging, and real-time FRB searches are addressed (progress in these areas is described by \citet{2024PASA...41...42S,2025arXiv250515204D}).

\section{Summary and conclusions}
\label{sec_summary}

Once more, standalone SKA-Low stations and EDA2 in particular turned out to be very flexible and productive scientific instruments for pulsar and transient sky monitoring. This work was possible thanks to the flexibility and quick turn around of the data as observations were typically processed and inspected within 24 hours from acquisition. 

In this paper, we have presented a new single pulse search pipeline implemented on a single station beam from EDA2 station. The effort was driven by searches for single pulses from FRBs, RRATs, magnetars and pulsars. No pulses from known repeating or other FRBs were detected during $\sim$100\,hours of observations, which resulted in upper limit for FRB rate R$\le$2500\,FRBs/day/sky. 

During the observing campaign nearly 100,000 sample of giant pulses (GPs) from the pulsar B0531+21 (Crab) were detected, which is to our knowledge the largest GP sample at frequencies $\le$300\,MHz. In this paper, we presented statistical analysis of this unprecedented sample of GPs. \red{On the timescales accessible to our analysis ($\gtrsim$1\,ms), the observed distribution of pulse arrival times is consistent with exponential distribution expected for purely random, Poisson process. These findings are consistent with the earlier observations that Crab GPs, unlike FRBs, do not show any signs of clustering observed in FRBs. Due to limitations of low-frequency observations, further high time resolution, high-frequency studies will be required to confirm the clustering of micro-bursts reported by \citet{2023ApJ...959..111L}. Our findings and the observed distribution of spectral luminosities do not exclude a possibility that extra-galactic Crab-like pulsars can produce weak repeating FRBs, but are unlikely progenitors of the entire population of FRBs. Especially, that given the observed distribution of spectral luminosities the wait times for typically bright FRBs from extra-galactic Crab-like pulsars would significantly exceed Hubble time.} Hence, the FRB rates would be many orders of magnitude lower than the observed rates (depending on the fluence, from 100s to 1000s per day per sky). 

The observed fluence distribution of bright (SNR$\ge$10) GPs can be well characterised by a single power law with the index $\alpha= -3.17\pm0.02$, and indexes $-3.13\pm0.02$ and $-3.59\pm0.06$ for distributions of GPs from main pulse and interpulse respectively. We observed a strong correlation (coefficient $\approx$0.9) of number of detected GPs with the measured scattering times, and when scattering is low ($\tau \sim $2\,ms) a much larger number of GPs was observed than at the time of high scattering ($\tau \sim$\,5\,ms). This is a result of fluence distribution normalisation being strongly modulated (correlation coefficient $\approx$0.9) by the scatter broadening time, which we recognise as the effect of refractive scintillation. The spectral index of scatter broadening was estimated to be $\beta = 3.6\pm0.1$, but the measurements were strongly affected by scattering conditions (including reduced SNR of GPs amongst other effects).

\red{We also observed indications of correlation of DM and scatter broadening time variations (correlation coefficient $\sim$0.7), which may be overestimated due to the impact of scatter broadening on DM measurements.}. Assuming that the event was caused by refractive scintillation, the estimated distance to plasma screen is of the order 1830\,pc which places the screen in the Perseus arm of the Galaxy or even the Crab Nebula itself. The velocity of the scintillation pattern, $V_{iss} \sim 2500\pm100$\,km/s, is similar to the expansion velocities observed in the Crab Nebula, which further supports this possibility. Modelling of DM and $\tau$ variations leads to size of plasma screen $\sim 10^{-5}$\,pc with Gaussian distributed cloud of electron plasma of mean electron density $\sim 400$e$^{-}$cm$^{-3}$, and peak electron density $\sim 2600$e$^{-}$cm$^{-3}$. This range of parameters is in agreement with earlier measurements, and matches electron densities in the filamentary structures of the Crab Nebula. 

Our detections of Crab GPs (the faintest down to 500\,Jy\,ms)  clearly demonstrate that even a single SKA-Low station can be a powerful instrument in searching for and studying pulsars, FRBs and other fast transients. Especially, once multi-beaming capability is commissioned on the newly built SKA-Low stations (4 beams planned in Array Assembly 1 and more later). Furthermore, as discussed in \citet{2024PASA...41...11S,2022aapr.confE...1S,2024PASA...41...42S} the high time resolution (millisecond) imaging of the entire visible hemisphere, leveraged with the real-time FRB search pipelines (e.g. \citet{2025arXiv250515204D}) can yield even one FRB detection per day. The presented analysis demonstrates the first single-beam attempt and proof of principle for such efforts on a larger scale. Once multiple beams or thousands of image pixels with the same sensitivity can be formed, the scientific output from a single station can be boosted by orders of magnitude. 

Our results provide a very optimistic prognostic for SKA-Low transient science, and of what may be possible in just a few years when hundreds of SKA-Low stations become operational. They also demonstrate SKA-Low potential to study local environments of pulsars, FRBs or other fast transients using low frequencies where propagation effects are much more pronounced than at higher frequencies. Given that nearby FRBs (z<0.1) can be similarly bright, and their DMs similar or only a few times higher than Crab pulsar's (e.g. FRB 20200120E has DM=87.82\,pc\,cm$^{-3}$ \citet{2021ApJ...910L..18B}), we conclude that the huge potential of SKA-Low for wide-field and all-sky searches for FRBs and transients is only a few steps away! 

% ~/Desktop/SKA/papers/2024/EDA2_FRBs/submission/REVIEW/LOG_NORMAL/images/FINAL/Python_Log10_Wait_Times_PAPER_GIMPED.png
% page 3 in /home/msok/Desktop/EDA2/papers/2024/EDA2_FRBs/20250612_distrib_time_interval_between_pulses_LOG-NORMAL_DISTR_PAPER.odt

\paragraph{Acknowledgments}
EDA2 is hosted by the MWA under an agreement via the MWA External Instruments Policy.
This scientific work makes use of the Murchison Radio-astronomy Observatory, operated by CSIRO. We acknowledge the Wajarri Yamatji people as the traditional owners of the Observatory site. This work was further supported by resources provided by the Pawsey Supercomputing Centre with funding from the Australian Government and the Government of Western Australia. The acquisition system was designed and purchased by INAF/Oxford University and the RX chain was design by INAF, as part of the SKA design and prototyping program. SD acknowledges the funding received from the ICRAR Summer Student program, which made parts of this work possible. We thank the anonymous referee for their helpful comments that significantly improved the quality of the manuscript.

% \todo{TBD:} We acknowledge Google and Google AI for help with various searches and simple tasks (e.g. generation of Python code for solution of cubic equation).

% \paragraph{Funding Statement}
% This research was supported by grants from the <funder-name><doi>(<award ID>); <funder-name><doi>(<award ID>).

% \paragraph{Competing Interests}
% A statement about any financial, professional, contractual or personal relationships or situations that could be perceived to impact the presentation of the work --- or `None' if none exist

% \paragraph{Data Availability Statement}
% A statement about how to access data, code and other materials allowing users to understand, verify and replicate findings --- e.g. Replication data and code can be found in Harvard Dataverse: \verb+\url{https://doi.org/link}+.

% \paragraph{Ethical Standards}
% The research meets all ethical guidelines, including adherence to the legal requirements of the study country.

\paragraph{Data availability}

The data can be made available on a reasonable request in their original or final formats. The original data were complex voltages from X and Y polarisations of a station beam in 1.08\,usec time resolutions. 

% \paragraph{Author Contributions}
% Please provide an author contributions statement using the CRediT taxonomy roles as a guide {\verb+\url{https://www.casrai.org/credit.html}+}. Conceptualization: A.A; A.B. Methodology: A.A; A.B. Data curation: A.C. Data visualisation: A.C. Writing original draft: A.A; A.B. All authors approved the final submitted draft.

%\endnote in some journals will behave like \footnote; and \printendnotes will not output anything. 
\printendnotes

% \defbibnote{preamble}{By default, this template uses \texttt{biblatex} and adopts the Chicago referencing style. However, the journal you’re submitting to may require a different reference style; specify the journal you're using with the class' \texttt{journal} option --- see lines 1--8 of \emph{sample.tex} for a list of options and instructions for selecting the journal.}

% \printbibliography[prenote={preamble}]
\printbibliography

\appendix
\section{Modelling pulse scatter broadening}
\label{appendix_one}

Pulse scatter broadening was modelled with a Gaussian (representing the on-set and peak of the pulse) with an exponential tail (only after the time of the pulse peak). Hence, it was mathematically described as:

\begin{equation}
f(t) = 
\begin{cases}
 G(t)\text{ for t }\le t_{peak} \\
 G(t) \cdot e^{-\frac{(t-t_{peak})}{\tau}} \text{ for t }\ge t_{peak} \\
\end{cases}
\label{eq_pulse_scatter_broadening}
\end{equation}

where $G(t)$ is the Gaussian profile used to describe the pulse peak:

\begin{equation}
G(t) = \frac{f_p}{\sqrt{2\pi \sigma_t}} \cdot e^{-\frac{(t-t_{peak})^2}{2\sigma_t^2}},
\end{equation}

and t is the time, $\sigma_t$ is standard deviation of the Gaussian profile, $t_{peak}$ is the time of the peak flux density $f_p$, and $\tau$ is the scatter broadening time. An example of a Crab giant pulse fitted with the above equations is shown in the right panel of Figure~\ref{fig_example_max_snr_gp}. 

We also tested more standard description where the Gaussian was convolved with an exponential function for times t above $t_{peak}$:

\begin{equation}
f(t) = 
\begin{cases}
 G(t)\text{ for t }\le t_{peak} \\
 \int_{-\infty}^{\infty}G(t') \cdot e^{-\frac{(t-t')}{\tau}} dt' \text{ for t }\ge t_{peak}.\\
\end{cases}
\label{eq_pulse_scatter_broadening_convol}
\end{equation}

However, both parametrisations yielded very similar results.

% As discussed in Section~\ref{subsec_modelling_dm}, the DM variations as a function of time were modelled as a ``blob of plasma'' passing through the line of sight between Earth and pulsar. The data were fitted with equation~\ref{} and the results of the fit are shown in Figure~\ref{}.

\section{DM verification by maximising peak flux of GPs}
\label{appendix_two}

\begin{figure}[hbt!]
\centering
% /home/msok/Desktop/EDA2/papers/2024/EDA2_FRBs/20250516_reprocess_data_to_get_DM_from_peak_PAPER.odt
% ~/Desktop/SKA/papers/2024/EDA2_FRBs/PAPER/DM_TIMING_vs_GPs/images/FINAL/DM_timing_vs_peakflux_vs_jb.png
\includegraphics[width=0.98\linewidth]{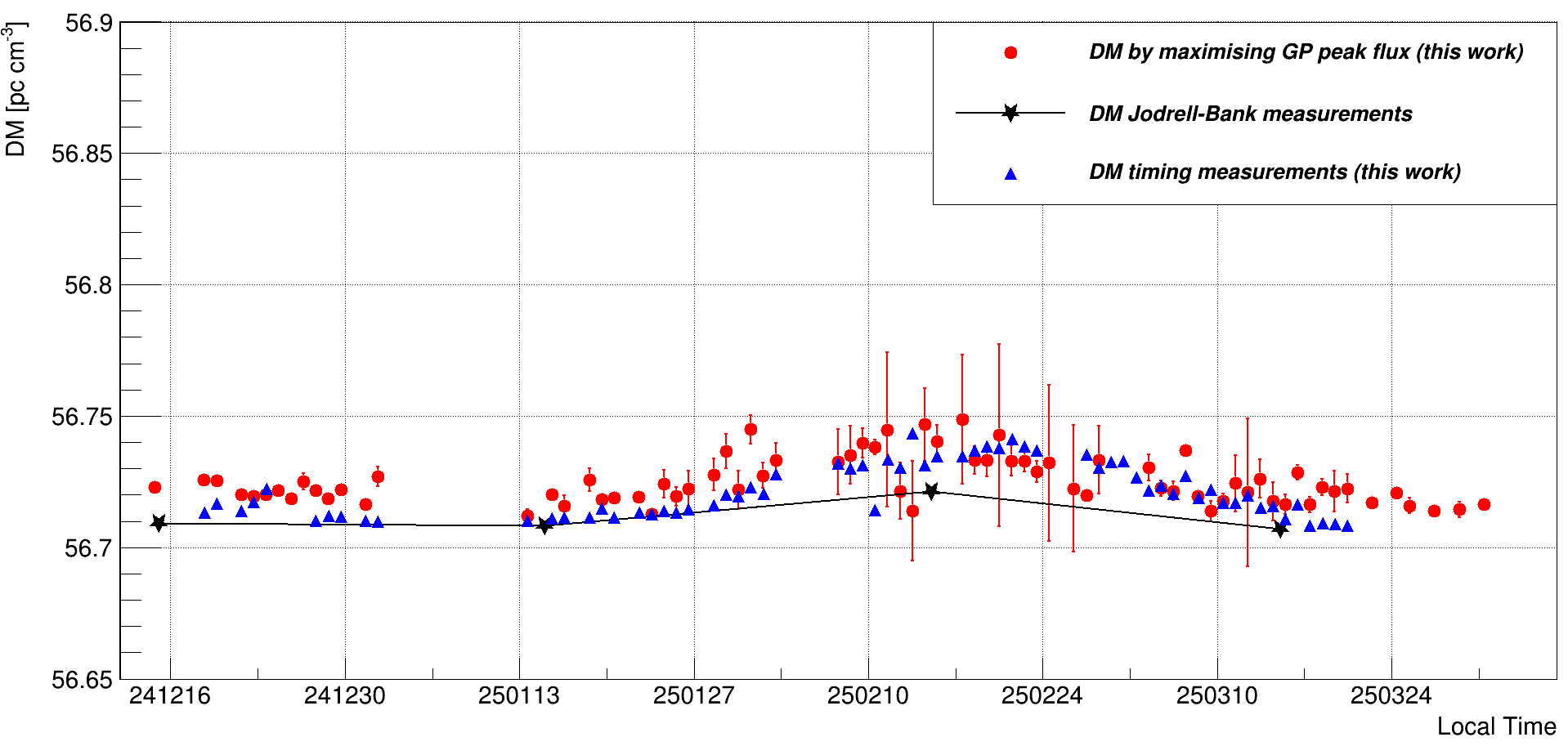}
\caption{The comparison of DM measured in timing analysis (blue triangles are the same data as in Figure~\ref{fit_dm_and_scat_vs_time}) and by maximising peak flux density of individual GPs (red filled circles). The black stars are Jodrell Bank measurements.}
\label{fig_dm_timing_vs_gps}
\end{figure}

It was noticed that high scattering may impact accuracy of timing-based DM measurements (mainly by shifting and changing shape of the average pulse profiles). Therefore, in order to verify the DM measurements obtained from timing we also measured DM using very bright (SNR$\ge$20) individual GPs. We tested the following methods of finding the optimal DM by : (a) maximising the slope of the leading edge between \sfrac{1}{4} and \sfrac{1}{2} of the peak, (b) maximising the slope of the entire leading edge, (c) maximising the peak flux density (i.e. SNR) and (d) maximising the rise time (difference between the peak time and start of the leading edge). The first two methods (a,b) were significantly affected by the scatter-broadening which led to large errors and variations of the resulting DM during high-scattering period. Therefore, we rely on method (c), which is also the typically used by single pulse search software packages finding DM maximising SNR/peak flux. The method (d) gives very similar results to method (c), but was also less reliable during high-scattering times. The comparison of DM measured with standard pulsar timing and by maximising peak flux density (method (c)) is shown in Figure~\ref{fig_dm_timing_vs_gps}. The errors in method (c) are much larger than from timing analysis during high scattering (between 10 and 28 February 2025) mainly because method (c) relies on single GPs (highly scattered and low-SNR at times), while timing analysis uses entire (1-hour) datasets. However, despite signs of systematic offset (especially between our measurements and Jodrell Bank data) the two methods are consistent, and indicate that the increase in DM, and its positive correlation with scattering time are real effects (i.e. not artefacts of analysis caused by the scatter broadening). This comparison, as well as comparison with DM measured by Jodrell Bank gave us confidence that our DM measurements are to within a factor of 2 accurate. 

\end{document}